\documentclass[letterpaper,titlepage,11pt]{article}
\pdfoutput=1

\usepackage{cite}
\usepackage{amsmath,amssymb,amsthm,mathrsfs,bbm}
\usepackage{latexsym,amscd,amsbsy,amsfonts,dsfont}
\usepackage{graphicx}
\usepackage{float}

\usepackage[
      colorlinks=true,
      linkcolor=blue,
      urlcolor=blue,
      filecolor=black,
      citecolor=blue,
      pdfstartview=FitV,
      pdftitle={},
        pdfauthor={Ryotaku Suzuki},
        pdfsubject={},
        pdfkeywords={},
        pdfpagemode=None,
        bookmarksopen=true
      ]{hyperref}
 \usepackage{caption}

\def\clock{{\count0=\time
           \divide\count0 60
           \ifnum\count0<10 0\fi\the\count0
           \multiply\count0 -60 \advance\count0 \time
           :\ifnum\count0<10 0\fi \the\count0
         }}
\newcommand{\timestamp}{{\small\vbox{\hbox{\tt\jobname.tex}
\hbox{\the\day/\the\month/\the\year, \clock}}}}

\newcommand{\cR}{\mathcal{R}} 

\newcommand{\cS}{{\mathcal S}}

\newcommand{\cJ}{{\mathcal J}}
\newcommand{\cM}{{\mathcal M}}
\newcommand{\cC}{{\mathcal C}}

\newcommand{\veps}{\varepsilon}
\newcommand{\ord}[1]{{\mathcal O}\left(#1\right)}
\newcommand{\ords}[1]{{\mathcal O}(#1)}
\newcommand{\fr}[1]{\frac{1}{#1}}
\newcommand{\nonum}{\nonumber\\ }
\newcommand{\hgfunc}[2]{ { \, {}_{#1}  F  {}_{#2} } }
\newcommand{\cout}[1]{}

\newcommand{\sbar}{{\rm \mathchar`-}}

\usepackage{bm}
\usepackage{latexsym}
\usepackage[latin1]{inputenc}
\usepackage{amsfonts}
\usepackage{amssymb}
\usepackage{amsmath}
\usepackage{subfigure}

\numberwithin{equation}{section}

\begin{document}

\begin{titlepage}

\vglue 2cm
\centerline{\LARGE \bf Black hole interactions at large $D$: brane blobology}

\vskip 1.6 cm
\centerline{\bf Ryotaku Suzuki$^{a,b}$}
\vskip 0.5cm

\centerline{\sl $^{a}$Departament de F{\'\i}sica Qu\`antica i Astrof\'{\i}sica, Institut de
Ci\`encies del Cosmos,}
\centerline{\sl  Universitat de
Barcelona, Mart\'{\i} i Franqu\`es 1, E-08028 Barcelona, Spain}
\centerline{\sl $^{b}$Department of Physics, Osaka City University,}
\centerline{\sl Sugimoto 3-3-138, Osaka 558-8585, Japan}

\smallskip
\vskip 0.5cm
\centerline{\small\tt s.ryotaku@icc.ub.edu}
\vskip 1cm

\centerline{\bf Abstract} \vskip 0.2cm \noindent
\noindent
In the large dimension ($D$) limit, Einstein's equation reduces to an effective theory on the horizon surface, drastically simplifying the black hole analysis.
Especially, the effective theory on the black brane has been successful in describing the non-linear dynamics not only of black branes, but also of compact black objects which are encoded
as solitary Gaussian-shaped lumps, {\it blobs}. For a rigidly rotating ansatz, in addition to axisymmetric deformed branches, various non-axisymmetric solutions have been found, such as black bars, which only stay stationary in the large $D$ limit. 

In this article, we demonstrate the blob approximation has a wider range of applicability
by formulating the interaction between blobs and subsequent dynamics.
We identify that this interaction occurs via thin necks connecting blobs.
Especially, black strings are well captured in this approximation sufficiently away from the perturbative regime. 
Highly deformed black dumbbells and ripples are also found to be tractable in the approximation.
By defining the local quantities, the effective force acting on distant blobs are evaluated as well.
These results reveal that the large $D$ effective theory is capable of describing not only individual black holes but also the gravitational interactions between them, as a full dynamical theory of interactive blobs, which we call {\it brane blobology}.

\end{titlepage}
\pagestyle{empty}
\small

\addtocontents{toc}{\protect\setcounter{tocdepth}{2}}
{
	\hypersetup{linkcolor=black,linktoc=all}
	\tableofcontents
}
\normalsize
\newpage

\pagestyle{plain}
\setcounter{page}{1}

\section{Introduction}
Black holes in higher dimensions exhibit colorful dynamics, in which horizons are stretched, bent and even pinched off~\cite{Emparan:2008eg}.
Remarkably, elongated horizons generically exhibit a long wave length instability, the Gregory-Laflamme (GL) type instability~\cite{Gregory:1993vy,Gregory:1994bj}, which can evolve to horizon pinch-off accompanied by the appearance of a naked curvature singularity illustrating a generic violation of strong cosmic censorship~\cite{Lehner:2010pn,Figueras:2015hkb,Bantilan:2019bvf}.
On the other hand, the same dynamics produces numerous stationary deformed solution branches coming out from the onset of the instability.

The large spacetime dimension limit, or the large $D$ limit~\cite{Emparan:2013moa,Emparan:2020inr} is an efficient approach to investigate such distinctive features of higher dimensional gravity.
In the large $D$ limit, black hole dynamics splits into two sectors with separate scales about the horizon scale $r_0$:
\begin{description}
\item[Decoupled sector] `Slow' dynamics with frequencies of $\ord{1/r_0}$, which is confined within a thin layer near the horizon, and contains GL-like self-gravitational deformations.
\item[Non-decoupled sector]
`Fast' dynamics with the gradient of $\ord{D/r_0}$, which corresponds to the radiative degrees of freedom in the time domain, as well as strong or short-scale horizon deformation in the space domain. 
\end{description}
Due to its non-perturbative-in-$1/D$ nature, the common framework has not been established for the fast dynamics.\footnote{Though, recent studies~\cite{Rozali:2018yrv,Emparan:2019obu} sheded some light on this sector. }
On the other hand, the slow dynamics is successfully reformulated into an effective theory of collective degrees of freedom on the horizon surface which is perturbative in $1/D$~\cite{Emparan:2015hwa, Emparan:2015gva,Emparan:2016sjk,Bhattacharyya:2015dva,
Bhattacharyya:2015fdk}. So far, the large $D$ effective theory approach has been applied to solve various black hole problems with or without Maxwell charge, cosmological constant, and even Gauss-Bonnet correction~\cite{Suzuki:2015axa,Herzog:2016hob,Rozali:2016yhw,Chen:2016fuy,Chen:2017rxa,Rozali:2017bll,Emparan:2018bmi,Casalderrey-Solana:2018uag,Iizuka:2018zgt,Tanabe:2015hda,Tanabe:2016pjr,Tanabe:2016opw,Chen:2017wpf,Mandlik:2018wnw,Chen:2018vbv}.

The large $D$ effective theory of black branes also admits Gaussian-shaped solitary lumps\footnote{Blobs are almost solitons except that the collision does not preserve the number of blobs due to the thermalization by the viscosity. }, or {\it blobs},
which encode compact black objects on a planer black brane~\cite{Andrade:2018nsz}.
In the blob solution, the Gaussian tail in the asymptotic region is expected to be matched with the equatorial part of black objects.
In the rigidly rotating setup, this blob approximation has made it possible to find a large variety of stationary deformed black holes bifurcating from the Myers-Perry family, such as black bars, ripples, dumbbells and flowers~\cite{Andrade:2018nsz,Andrade:2018rcx,Licht:2020odx}.
It is also possible to have multiple blobs, where each blob travels almost independently until they collide to merge or split~\cite{Andrade:2018yqu,Andrade:2019edf,Andrade:2020ilm}. One should bear in mind that, in the multiple configuration, the leading order theory does not resolve whether or not the horizon is actually joined or split, since the thickness between blobs can be arbitrary thin. 
When $1/D$ corrections are included, the thin region leads to a breakdown of the $1/D$ expansion
 indicating disconnected horizons.

In the large $D$ limit, the gravitational force from a black hole is suppressed non-perturbatively in $1/D$ except in the thin near horizon region, and hence the interaction between black holes is typically negligible in the $1/D$-expansion.
The blobs, nevertheless, can weakly interact each other.
For example, in the collision between two blobs at large impact parameter, it has been observed that
the two blobs orbit several rounds before they collide or split away, as if they are attracted to each~\cite{Andrade:2020ilm}.
It has also been noticed that black dumbbells that are highly deformed
form a multiple binary of separate blobs, and black ripples resemble multiple concentric ring-shaped blobs~\cite{Licht:2020odx}. 
In all of these configurations, there are blobs connected by thin neck regions. Since the blobs are rotating, the centrifugal forces should be sustained by a certain attraction between blobs. 
These interactions can be understood by regarding multiple blobs as either of black holes whose horizons are almost touching (so that near horizons have some overlap) or large lumps on a single black object which are connected to each other by thin necks.
As mentioned above, to leading order at large $D$, it is not possible to strictly distinguish if two blobs connected by a thin neck are different black holes, or part of the same one.
 
\begin{figure}[t]
\begin{center}
\includegraphics[width=8cm]{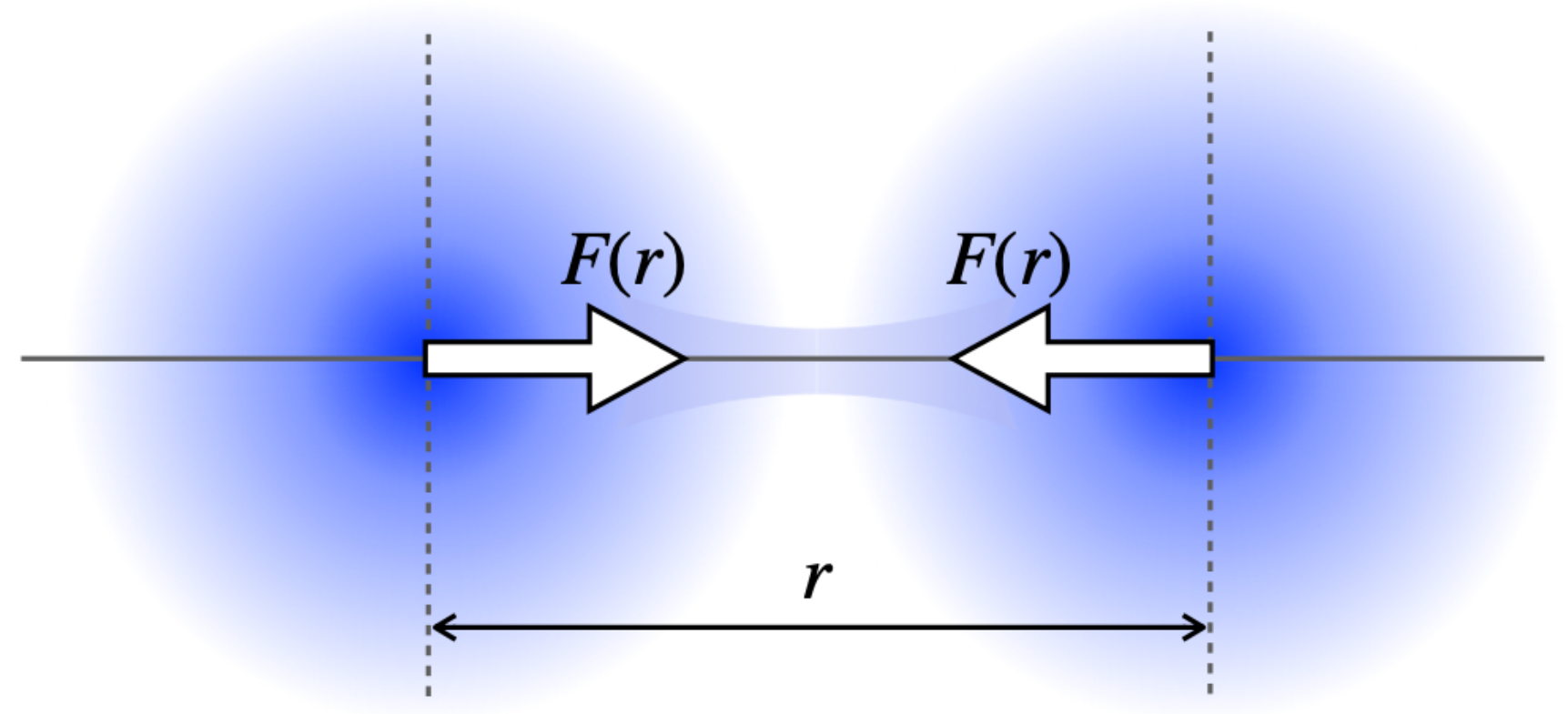}
\caption{Effective force between blobs in the large $D$ effective theory. }
\label{fig:blobattraction}
\end{center}
\end{figure}

In this article, we formulate analytical descriptions for the blob-blob interaction by resolving thin neck regions between blobs. The resulting formula is written as a trans-series with respect to the distance between blobs. Remarkably, we can approximate the effective attraction force between two blobs mediated by the thin neck (figure~\ref{fig:blobattraction})
\begin{equation}
 F(r) \simeq  \frac{\sqrt{\cM_A \cM_B}\,r^2}{2\sqrt{2\pi}}e^{-\frac{r^2}{8}},
\end{equation}
where $\cM_A$ and $\cM_B$ are the blob mass and $r$ is the distance between blobs.

This article is organized as follows.
In section~\ref{sec-bs}, we revisit the non-uniform black string analysis in the highly deformed regime by expanding in small tension. In sections~\ref{sec:dumb} and \ref{sec:ripple}, highly deformed black dumbbells and ripples are studied by using the same technique in the limit of zero rotation $\Omega \to 0$.
In section~\ref{sec:kinematic}, we formulate the effective kinematics of blobs by defining localized variables for each blob, and then derive the approximate equation of motion for blobs.
We summarize and discuss the possible extension of the result in section~\ref{sec:summary}.

\subsection*{Brane blobology}
The term 'brane blobology' was introduced in~\cite{Andrade:2020ilm}, as a pun on the {\it brain blobology} in the neuroscience. However, we do not claim any scientific relation.


\section{Large $D$ effective theory}\label{sec:review}
In this section, the large $D$ effective theory on the black brane is briefly reviewed~\cite{Emparan:2015hwa, Bhattacharyya:2015dva,Emparan:2015gva,
Bhattacharyya:2015fdk,Emparan:2016sjk, Bhattacharyya:2016nhn, Dandekar:2017aiv}.
The effective theory of the $p$-brane is embedded in a $D=p+3+n$ background%
\footnote{Throughout this article, we focus on the vacuum, asymptotically-flat spacetime. The extension to the Einstein-Maxwell theory and inclusion of the cosmological constant will be discussed in section~\ref{sec:summary}.}
\begin{equation}
ds^2 = -dt^2 +dr^2+ \fr{n} dx^i dx_i + r^2 d\Omega_{n+1}^2 \label{eq:eft-ansatz}
\end{equation}
where $i=1,\dots,p$ and $d\Omega_{n+1}^2$ is $S^{n+1}$-metric. Here, we set the physical horizon scale $r_0=1$. Assuming $n\simeq D$, it is convenient to expand by $1/n$ instead of $1/D$.
The $1/n$ scaling for $x$ coordinates is introduced  to capture the nonlinear dynamics along those directions.
In the large $D$ limit, the entire spacetime geometry is described in terms of the collective degrees of freedom : the mass density $m(t,x)$ and momentum density $p_i(t,x)$. 
By introducing the velocity field $p_i= m v_i+\partial_i m$,
the effective equation take the form of fluid equations,
\begin{align}
& \partial_t m + \nabla_i (mv^i) = 0,\quad
 \partial_t (mv^i) + \nabla_j (mv^i v^j+\tau^{ij} ) = 0,\label{eq:eft-effectiveLO}\\
&\qquad \tau_{ij} := -m\delta_{ij}-2m\nabla_{(i}v_{j)} - m\nabla_i \nabla_j \log m\nonumber
\end{align}
where the indices are raised and lowered with the flat spacial metric $\delta_{ij}$ and $\nabla_i$ is the covariant derivative for $\delta_{ij}$. 
The quasi-local stress tensor for the $p$-brane is given by
\begin{align}
&T^{tt} = m,\quad T^{ti} = m v^i,\quad T^{ij} =  m v^i v^j + \tau^{ij},\nonum
&\qquad\tau_{ij} = -2m\partial_{(i} v_{j)}- m\delta_{ij} - m \partial_i \partial_j \log m,\label{eq:eft-stresstensor}
\end{align}
where the actual dimensionful stress tensor is given by 
\begin{equation}
{\bf T}_{\mu\nu} := \frac{(n+1)\Omega_{n+1}}{16\pi G} T_{\mu\nu}.
\end{equation}
The effective equation~(\ref{eq:eft-effectiveLO}) is equivalent to the conservation of this tensor
\begin{eqnarray}
\partial_\mu T^{\mu\nu} = 0.
\end{eqnarray}
The physical quantities of the black brane are given by integrating the stress tensor%
\footnote{For simplicity, the spacial scaling $1/n$ in eq.~(\ref{eq:eft-ansatz}) is omitted from the volume integral}
\begin{equation}
 {\cal M} = \int_{{\mathbb R}^p} T^{tt} d^px ,\quad {\cal P}_i = \int_{{\mathbb R}^p} T^{ti} d^px ,\quad {\cal J}_{ij} = \int_{{\mathbb R}^p} (x_i T^{t}{}_j-x_j T^t{}_j) d^px.
\end{equation}

\paragraph{Gaussian blobs}
Besides ordinary brane solutions, this equation exhibits stable Gaussian solutions, or {\it blobs}. For $p=2$, the general rotating blob solution is given by
\begin{align}
 m(t,z) = m_0\exp\left(-\frac{(x-ut-b)^2}{2(1+a^2)}\right),\quad v^i(t,z) = u^i + \frac{a}{1+a^2}\veps_{ij}(x^j-u^j t).
\end{align}
where $a$ corresponds to the spin parameter of the singly spinning Myers-Perry black hole, $u^i$ and $b^i$ are arbitrary constants reflecting the boost invariance and Galilean symmetry of the system. In this article, these strict Gaussian solutions are called as {\it Gaussian blobs} apart from other deformed blob solutions. For general blob solutions, we only assume regularity and Gaussian-type fall off in the asymptotic region so that the physical quantities remains finite.

\paragraph{Rigidly rotating solutions}
With the rigidly rotating ansatz
\begin{equation}
( \partial_t + v^i \partial_i) m = 0,\quad \partial_t v^i = 0 ,\quad \nabla_{(i}v_{j)}=0,
\end{equation}
the effective equation is integrated to the soap bubble equation~\cite{Andrade:2018nsz}
\begin{equation}
\nabla^2 \cR + \fr{2} (\nabla \cR)^2 +\cR+\fr{2}v^2= C,\label{eq:eft-effective-rigid}
\end{equation}
where $C$ is an integration constant fixing the mass scale
and $\cR = \log m$ denotes the horizon radius as\footnote{The event horizon and apparent horizon are degenerately equal to $m^\fr{D}$ in the leading order.}
\begin{equation}
r_H \simeq m^\fr{D} \simeq 1 + \fr{D} \cR.
\end{equation}


\section{Highly deformed non-uniform black strings}\label{sec-bs}
Let us start by revisiting the black string analysis in the $1+1$ effective theory~\cite{Emparan:2015hwa,Suzuki:2015axa,Emparan:2015gva,Emparan:2018bmi}.
The effective equation for the static black string is given by
\begin{equation}
\cR''(z) + \fr{2} \cR'(z)^2 + \cR(z)  = C,\label{eq:nubs-effeq}
\end{equation}
where the constant $C$ determines the mass scale.
The spacetime is compactified in the $z$ coordinate with a period $L$.
Multiplying by $e^{\cR(z)}\cR'(z)$, this can be integrated to
\begin{equation}
\left(\fr{2}\cR'^2 + \cR-1-C\right) e^{\cR} = -\tau,\label{eq:nubs-effeq-tau}
\end{equation} 
where $\tau$ is the string tension, which coincides with $\tau_{zz}$ in the stress tensor~(\ref{eq:eft-stresstensor}). The lower tension leads to higher non-uniformity of the solution. Without loss of generality, we fix $C=0$ in the following analysis, in which the non-uniform black strings are obtained for $0<\tau<1$, and the uniform string at the bifurcation point for $\tau = 1$. 
At the zero tension limit $\tau\to 0$, the period $L$ goes to infinity and the solution goes to the strict Gaussian blob
\begin{equation}
 \cR(z) = 1-\frac{z^2}{2}.\label{eq:nubs-gaussian-0}
\end{equation}
The mass of black string is defined by the integration over the period,
\begin{equation}
 {\cal M} := \int_{-\frac{L}{2}}^{\frac{L}{2}} e^{\cR(z)} dz.
\end{equation}
In particular, the Gaussian blob mass is given by the Gaussian integral
\begin{equation}
 \cM_0  := \int_{-\infty}^\infty e^{1-\frac{z^2}{2}}dz = \sqrt{2\pi} \, e. \label{eq:nubs-gblob-mass}
\end{equation}
Once the solution $(\cR(z;\tau),\tau)$ is obtained, the mass scale is easily recovered by
\begin{equation}
\cR(z;\tau) \to \cR_C(z;\tau_C) = \cR(z; e^{-C}\tau_C) + C,\quad {\cal M} \to {\cal M}_C = e^C{\cal M},\quad \tau \to \tau_C = e^{C}\tau.
\end{equation}

\subsection{Gaussian blob with small tension}
Assuming $\tau \ll 1$, the solution should be expanded from the zero tension Gaussian blob
\begin{equation}
\cR(z) = 1-\fr{2}z^2 +\tau\, \delta \cR(z).\label{eq:nubs-gaussian-dR}
\end{equation}
The linear correction is solved by
\begin{equation}
\delta \cR(z) = - e^{-1+\frac{z^2}{2}}+e^{-1}\sqrt{\frac{\pi}{2}}\,z\, {\rm erfi}(z/\sqrt{2}).
\label{eq:nubs-gaussian-1}
\end{equation}
In the mass profile, this correction falls off much slower than the Gaussian at large $z$
\begin{equation}
m(z) = e^{\cR(z)} \simeq e^{1-\frac{z^2}{2}} + \frac{\tau}{z^2},
\end{equation}
where the perturbation breaks down.
However, we will see the solution is smoothly continued to the neck waist solution where these two terms become comparable.

\subsection{Neck waist}\label{sec:nubs-neck-waist}
\begin{figure}[t]
\begin{center}
\includegraphics[width=14cm]{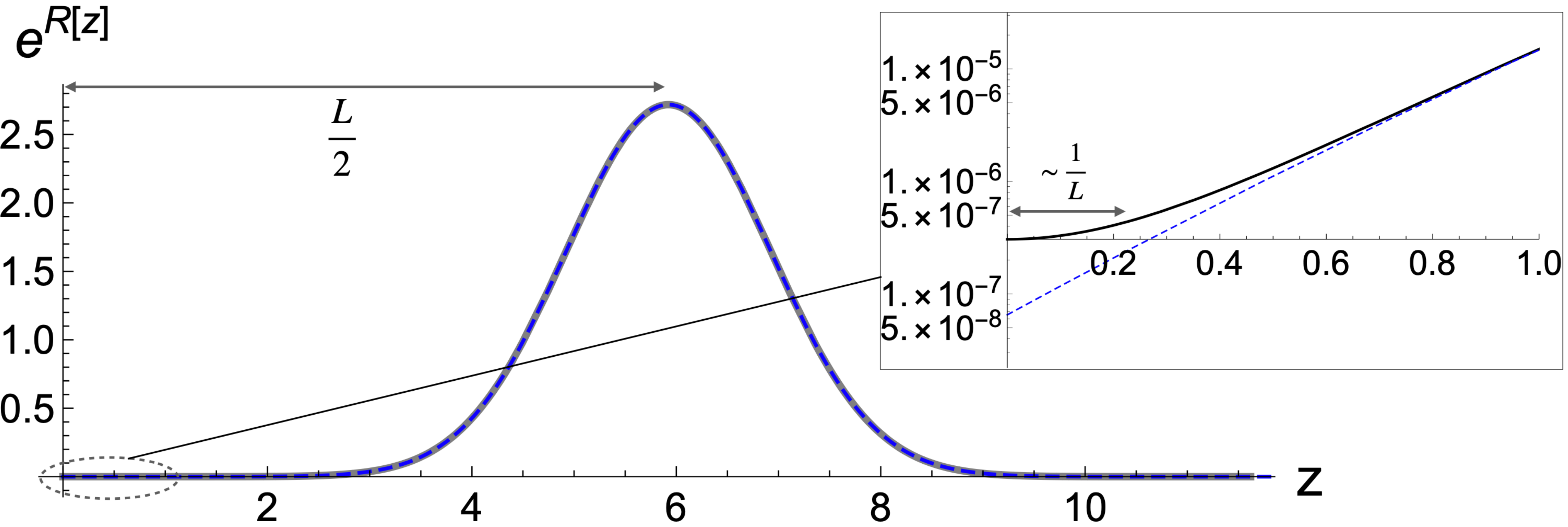}
\caption{A mass profile of the non-uniform black string with $\cR(0)=-15$. The Gaussian solution is plotted by the blue dashed curve.}
\label{fig:nubs-prof}
\end{center}
\end{figure}

Since the period $L$ becomes large at $\tau \to 0$, we study the neck region with the expansion from $L=\infty$.
By numerically solving the effective equation for the small tension, 
we observe that the long Gaussian tail which stretches as $L$ is smoothly connected to a short waist which shrinks as $L^{-1}$ (figure \ref{fig:nubs-prof}). The amplitude of the waist is also estimated by eq.~(\ref{eq:nubs-gaussian-0}) as $\cR \simeq -L^2/8$.
From this observation, we introduce another scaling on the neck waist
\begin{equation}
 \cR = -\fr{8} L^2+\bar{\cR}(\xi),\quad z = \pm\frac{L}{2}+ \frac{2\xi}{L}.
\end{equation}
Plugging this into the effective equation~(\ref{eq:nubs-effeq}), we obtain
\begin{equation}
   \bar{\cR}''(\xi) + \fr{2} \bar{\cR}'(\xi)^2- \fr{2} = -\frac{4}{L^2}\bar{\cR}(\xi).\label{eq:nubs-effeq-neck}
\end{equation}
This equation can be solved by expanding $\bar{\cR}(\xi)$ with $1/L^2$,
\begin{equation}
\bar{\cR}(\xi) = \sum_{i=0}^\infty L^{-2i}\bar{\cR}_i(\xi).
\end{equation}
Since eq.~(\ref{eq:nubs-effeq-neck}) only includes the derivative terms in the leading order,
we always have a translation degree of freedom at each order, which is used to set the minimum at $\xi=0$. This also sets the solution to be symmetric in $\xi \to -\xi$.
In the leading order, the solution becomes
\begin{equation}
\bar{\cR}_0(\xi) = \bar{r}_0 + 2 \log \cosh(\xi/2),\label{eq:neck-sol-0}
\end{equation}
where $ \bar{r}_0 $ is an integration constant which contributes to the minimum value at the neck. The leading order solution has the following asymptotic behavior at $\xi \to \pm \infty$,
\begin{equation}
\bar{\cR}_0(\xi) \simeq
\pm\xi+\bar{r}_0-2\log 2 + 2 e^{-|\xi|}.\label{eq:neck-sol-0-as}
\end{equation}
This gives the match with the Gaussian blobs~(\ref{eq:nubs-gaussian-0}) at both sides $z=\pm L/2$,
\begin{equation}
 -\frac{L^2}{8} + \bar{\cR}_0(\xi) +\ord{L^{-2}} \simeq - \fr{2}\left(\frac{L}{2}\mp\frac{2\xi}{L}\right)^2 +  \bar{r}_0  -2\log 2 +\ord{e^{-|\xi|},L^{-2}} ,
\end{equation}
where the matching region is given by $1\ll \xi \ll L$.
The match with the Gaussian solution~(\ref{eq:nubs-gaussian-0}) determines the integration constant
\begin{equation}
\bar{r}_0 = 1+2\log 2,\label{eq:nubs-match-r0}
\end{equation}
which gives the neck minimum up to $\ord{L^{0}}$,
\begin{equation}
  \cR_{\rm min} = -\frac{L^2}{8} + 1+2\log2+\ord{L^{-2}}.\label{eq:Rmin-L-LO}
\end{equation}
Since the tension is given by the minimum value, we can express the tension in terms of the compactification length,
\begin{equation}
\tau = (1-\cR_{\rm min}) e^{\cR_{\rm min}}  \simeq \fr{2} L^2 e^{1-\frac{L^2}{8}}.\label{eq:tension-L-LO}
\end{equation}
Inversely, the scale $L$ is expressed by the tension
\begin{equation}
 L \simeq 2\sqrt{-2W_{-1}(-\tau/(4e))},
\end{equation}
where $W_{-1}(x)$ is the lower branch of the Lambert W function.
At the limit $\tau \to 0$, the scale $L$ admits a logarithmic dependence to the tension,
\begin{equation}
 L \simeq 2\sqrt{-2\log \tau}.
\end{equation}
By repeating the procedure, we can improve the expansion in $1/L^2$, 
\begin{align}
&  \cR_{\rm min} = -\frac{L^2}{8} + 1+2\log2+\fr{L^2}\left( \frac{2\pi^2}{3} + 16 \log2\right) \nonum
  &+\fr{L^4}\left( \frac{64\pi^2}{3}+96 \zeta(3)+128 (\log^22+\log2)\right)+\ord{L^{-6}},\label{eq:Rmin-L-NLO}
\end{align}
\begin{equation}
 \tau = \frac{L^2}{2}e^{1-\frac{L^2}{8}}\left(1+\frac{2\pi^2}{3L^2}+\frac{16\pi^2+2\pi^4/9+96 \zeta(3)}{L^4}+\ord{L^{-6}}\right).
 \label{eq:tension-L-NLO}
\end{equation}
The detail of the calculation is given in Appendix~\ref{sec:neck-sub}.
In figure \ref{fig:nubs-L_Rmin}, we compare the tension with the numerical calculation, together with the perturbatively constructed result from the uniform solution.
\begin{figure}[t]
\begin{center}
\includegraphics[width=7cm]{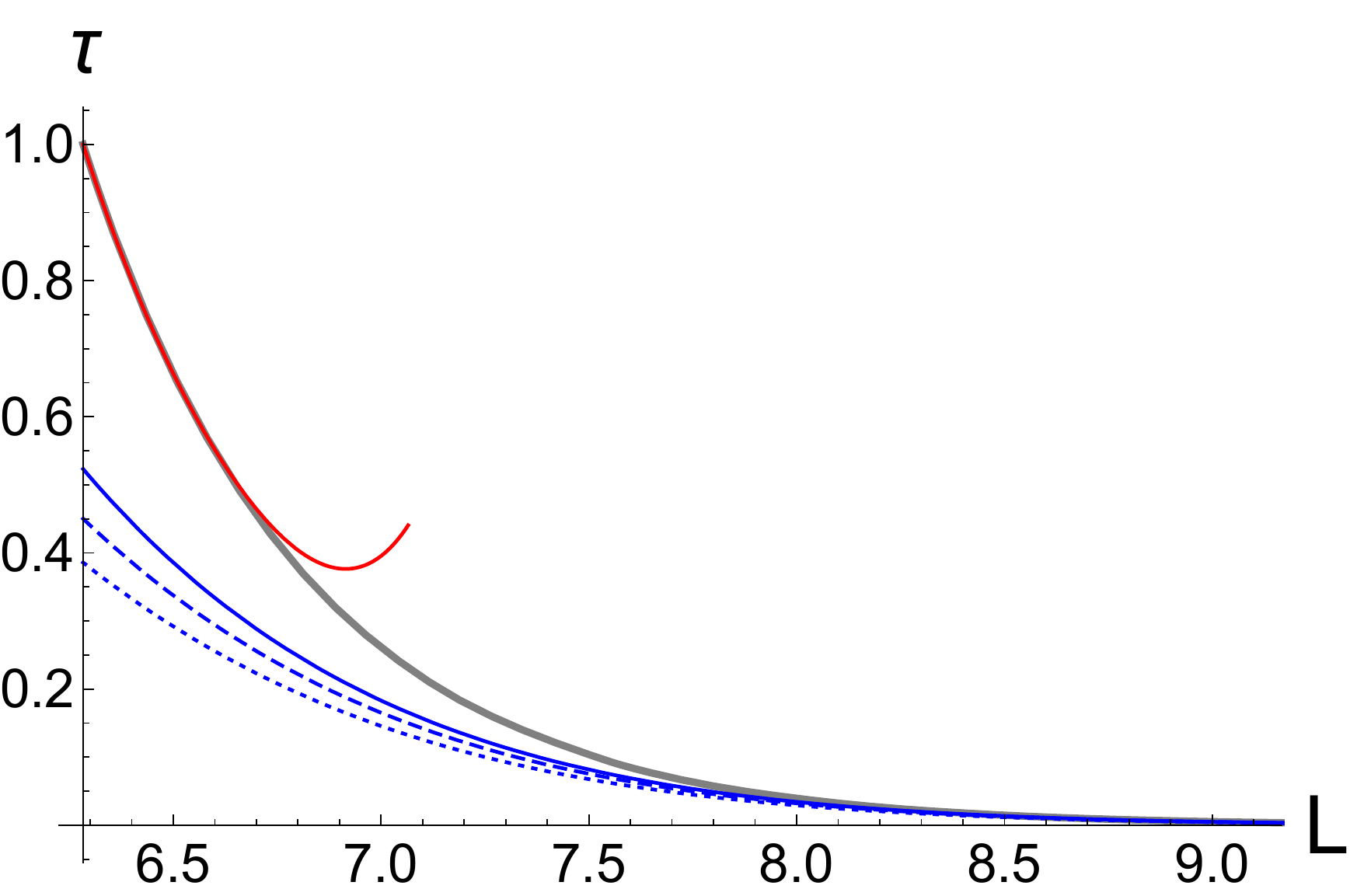}
\caption{The tension for given period $L$. The gray curve represents the numerical result. The red curve is given by the perturbative expansion in the small non-uniformity up to $8$-th order. 
The blue dotted, dashed and solid curves correspond the blob approximation up to LO, NLO and NNLO in eq.~(\ref{eq:tension-L-NLO}), respectively.}
\label{fig:nubs-L_Rmin}
\end{center}
\end{figure}

\paragraph{Match with $\ords{\tau}$-correction}
Given the relation between the tension and compactification scale, we can evaluate the behavior of the correction to the Gaussian blob~(\ref{eq:nubs-gaussian-1}) in the matching region. For $z\to \infty$, 
eq.~(\ref{eq:nubs-gaussian-1}) has the asymptotic behavior
\begin{equation}
 \delta \cR(z) = e^{-1}\sqrt{\frac{\pi}{2}} z\, {\rm erfi}(z/\sqrt{2})-e^{-1+\frac{z^2}{2}} \simeq \fr{z^2}e^{-1+\frac{z^2}{2}}(1+\ord{z^{-2}}).
\end{equation}
Expanding at $z = \pm L/2 + \xi/L$ with eq.~(\ref{eq:tension-L-LO}), 
the correction term is shown to be $\ord{L^0}$,
\begin{equation}
\tau \delta \cR \simeq 2 e^{-|\xi|} + \ord{\tau,L^{-2}}.
\end{equation}
This gives a consistent match with eq.~(\ref{eq:neck-sol-0-as}) at $\ord{e^{-|\xi|}}$.
Here, we should note that $\ords{\tau^k}$-correction gives $\tau^k \delta^{[k]} \cR \sim \tau^k e^\frac{kz^2}{2}\sim e^{-k|\xi|}$, resulting in the similar consistent match up to the relevant order in $1/L^2$.

\subsection{Phase diagram}
Finally, we evaluate the total mass. 
The total mass is obtained by summing both contributions from the blob and neck parts,
\begin{equation}
 {\cal M} = {\cal M}_{\rm blob} + {\cal M}_{\rm neck}.
\end{equation}
After the summation, the actual matching position should not appear in the result.
The contribution from the blob part is given by
\begin{align}
& {\cal M}_{\rm blob} = 2\int_0^{z_c} dz \left [e^{1-\frac{z^2}{2}}+\tau \left(\sqrt{\frac{\pi}{2}} z\, {\rm erfi}(z/\sqrt{2})-1\right)\right]\nonum
 &\qquad = \sqrt{2\pi} e - \sqrt{2\pi} e\, {\rm erfc}(z_c/\sqrt{2})
-\sqrt{2\pi} \tau e^{-\frac{z_c^2}{2}} {\rm erfi}(z_c/\sqrt{2}),\label{eq:nubs-mass-blob-exact}
\end{align}
where $z_c$ is a cut off of the integral.
We set $z=z_c$ at some point in the matching region
\begin{equation}
 z_c = \frac{L}{2} - \frac{2\xi_c}{L},\qquad (1\ll \xi_c \ll L).\label{eq:nubs-matching-region}
\end{equation}
The asymptotic expansion at large $L$ leads to
\begin{align}
& \frac{{\cal M}_{\rm blob}}{ \cM_0 } = 1 - \frac{2 L e^{-\frac{L^2}{8}}}{\sqrt{2\pi}} \left[1+\frac{2}{L^2}\left(\frac{\pi^2}{3}+2+2\xi_c+e^{\xi_c}\right)+\ord{\tau,e^{-\xi_c},L^{-4}}\right],
\end{align}
where eq.~(\ref{eq:tension-L-NLO}) is used and the mass is normalized by the Gaussian blob mass~(\ref{eq:nubs-gblob-mass}). We take the terms up to the first sub-leading order in $1/L^2$ to see the cut off effect. It turns out that $\ords{\tau^k}$-correction also has the contribution of $\ords{\tau^k e^\frac{(k-1)z_c^2}{2}} = \ord{\tau e^{(1-k)\xi_c}}$, which should cancel out with the neck terms. Here, we simply neglect $\ord{e^{-\xi_c}}$-terms, so that we do not have to care $\ord{\tau^k}$-corrections for $k>1$.
For the neck part, we take the solution up to LO in $1/L^2$,
\begin{align}
& \frac{{\cal M}_{\rm neck}}{ \cM_0 } = \frac{2}{ \cM_0 } \int_{z_c}^{L/2} dz e^{\cR(z)} = \frac{16}{\sqrt{2\pi} L}e^{-\frac{L^2}{8}} \int_0^{\xi_c} d\xi \cosh^2(\xi/2)\left(1+\ord{L^{-2}}\right)\nonum
&\hspace{1cm}=\frac{4e^{-\frac{L^2}{8}}}{\sqrt{2\pi} L} \left(2\xi_c + e^{\xi_c}+\ord{e^{-\xi_c},L^{-2}}\right).\label{eq:nubs-m-neck-LO}
\end{align}
One can see that the cut off contributions in both expansions cancel out in the total mass.
Expanding up to one higher order in $1/L^2$, the same cancellation occurs (see Appendix.\ref{sec:neck-sub-mass}) and the total mass becomes
\begin{equation}
\frac{{\cal M}}{ \cM_0 } = 1  - \frac{2 L e^{-\frac{L^2}{8}}}{\sqrt{2\pi}} \left[1+\frac{4}{L^2}\left(\frac{\pi^2}{6}+1\right)+\frac{2(432\zeta(3)+\pi^4+60\pi^2-72)}{9L^4}+\ord{L^{-6}}\right].\label{eq:bs-mass-res}
\end{equation}
\begin{figure}[t]
\begin{center}
\includegraphics[width=6cm]{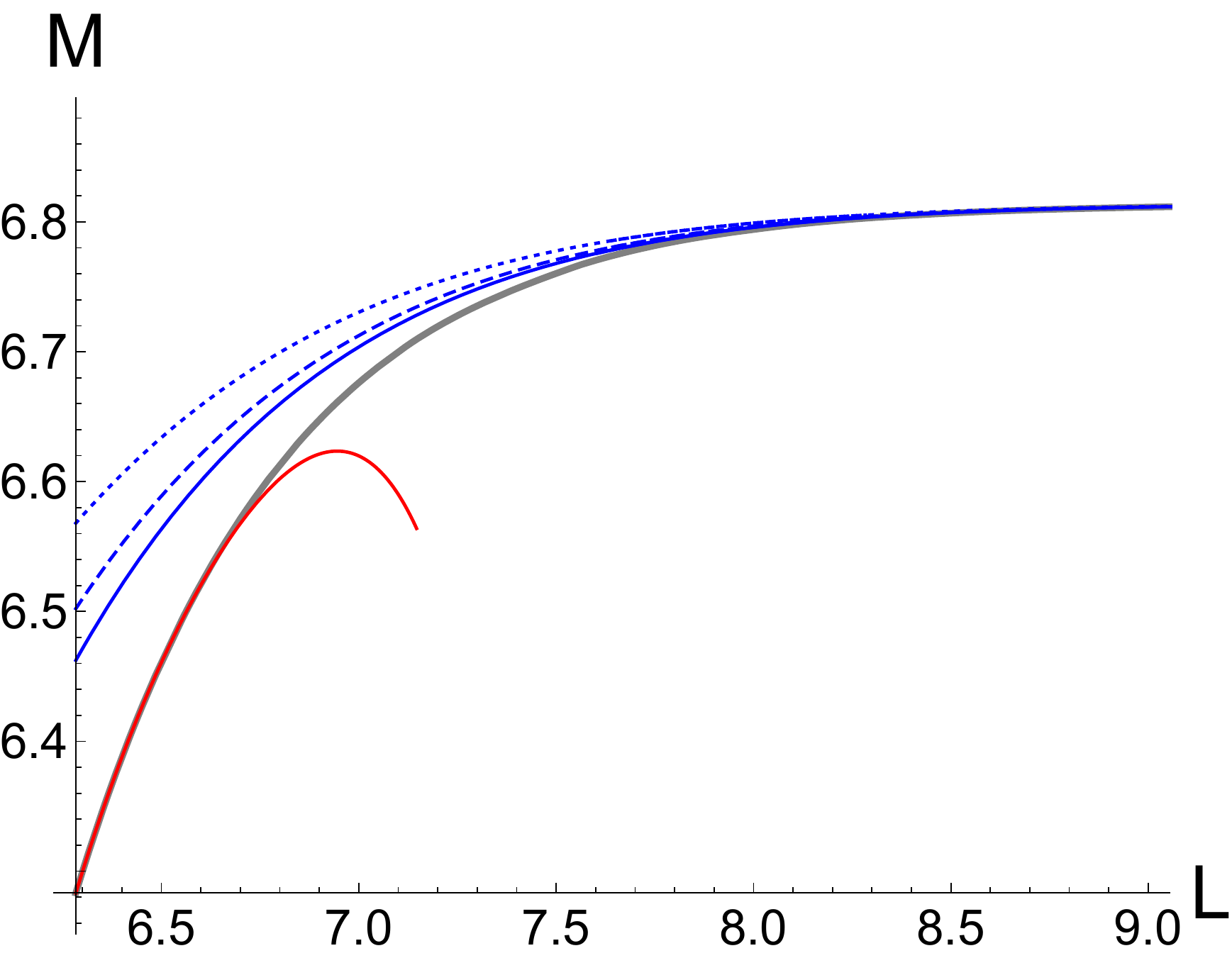}
\includegraphics[width=6cm]{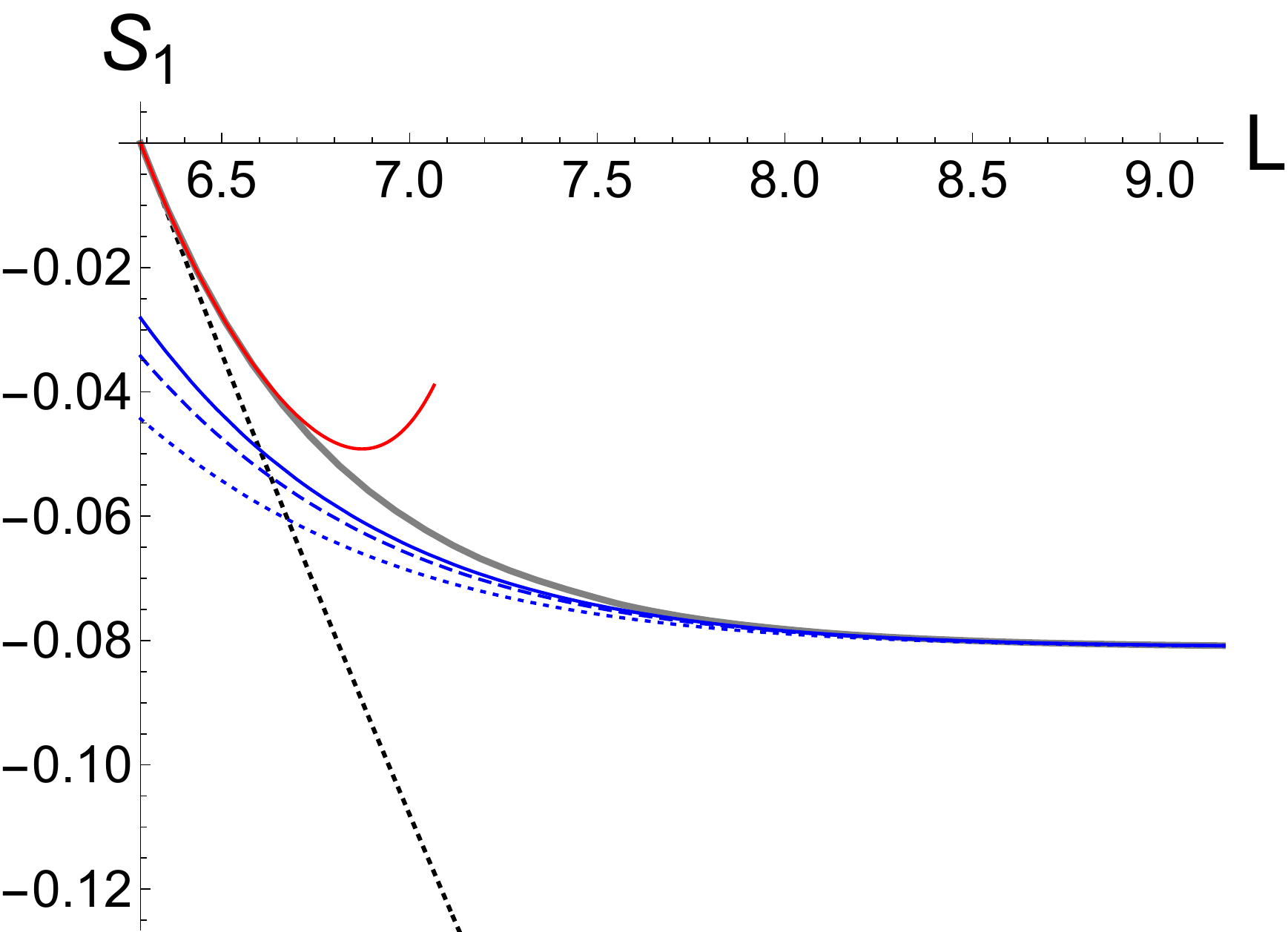}
\caption{The mass and the mass-normalized, scale-invariant entropy of the nonuniform black string for given period $L$. The gray and red curves correspond respectively to the numerical result and perturbative expansion in the small non-uniformity up to $8$-th order. The blue dotted, dashed and solid curves are the blob approximation up to LO, NLO and NNLO, respectively. The black dotted curve in the right panel is the scale-invariant entropy of the uniform string given by ${\cal S}_1 = \log (2\pi/L)$.}
\label{fig:nubs-M_L}
\end{center}
\end{figure}
Using the effective equation, the mass-normalized, scale-invariant entropy~\cite{Andrade:2020ilm} is
given by the total mass
\begin{equation}
  {\cal S}_1=  - \log {\cal M} + \log (2\pi).
\end{equation}
where the scaling is fixed so that the uniform black string with $L=2\pi$ gives zero entropy.
These results approximate well the numerical result in the highly deformed regime (figure~\ref{fig:nubs-M_L}). 

Up to now, we have only considered the linear order correction of the tension.
Including higher orders of the tension will result in the trans-series of $1/L^{2}$ and $e^{-\frac{L^2}{8}}$, which will require the knowledge of the trans-series expansion of error functions.
One can notice that, although the convergence of the $1/L^2$-expansion is not bad close to $L=2\pi$, the formula starts to disagree with the numerical result. This would also imply the presence of correction terms in the trans-series expansion.

Lastly, we consider the limit $L \sim \sqrt{D}$, which gives a neck amplitude of $\cR \sim -D $ and the $1/D$ expansion breaks down there.
In this regime, the physical compactification length reaches $\ord{1}$, in which one can observe the smooth topology-changing transition by taking the proper scaling on the neck~\cite{Emparan:2019obu}. On the other hand, eq.~(\ref{eq:bs-mass-res}) reduces to a trans-series expansion in $1/D$ and $e^{-\alpha D}$, which is consistent with the large $D$ conifold analysis in which the back reaction from the pinching region becomes non-perturbative in $1/D$.
This implies the trans-series expansion in the blob approximation captures the smooth transition from the decoupled to the non-decoupled sector in the topology-changing phase.
In ref.~\cite{Casalderrey-Solana:2018uag}, trans-series in the perturbative and non-perturbative expansion in $1/D$ was also obtained for non-decoupled quasinormal modes.


\section{Highly deformed black dumbbells}\label{sec:dumb}
The same analysis can be applied to the black dumbbells found in the $2+1$ rigidly rotating system~\cite{Licht:2020odx}.
Dumbbells are deformed branches bifurcating from the onsets of the instability on the black bar. 
In ref.~\cite{Licht:2020odx}, 
dumbbells in highly deformed phase resemble arrays of multiple Gaussian blobs almost separate from each other.
The blobs are aligned in almost equal distances logarithmically growing in $\Omega$, which would imply a pinch-off transition to the multiple spherical black holes.
We show that these behaviors are actually described by the interaction between blobs and necks.
\subsection{Setup}
The effective equation for the $2+1$ rigidly rotating black brane is given by~\cite{Andrade:2018nsz}
\begin{equation}
 \partial_r^2 \cR + \fr{r} \partial_r \cR + \fr{r^2} \partial_\phi^2 \cR +  \fr{2} \left((\partial_r \cR)^2+\fr{r^2}(\partial_\phi \cR)^2\right) + \cR + \fr{2}\Omega^2 r^2 = 0,\label{eq:dumb-2+1eq}
\end{equation}
where the solution is rigidly rotating as $\cR(t,r,\phi) = \cR(r,\phi-\Omega \, t)$.
Black bars and dumbbells are solved by introducing the co-rotating Cartesian coordinate,
\begin{equation}
 x := r \cos(\phi-\Omega t),\quad y := r \sin(\phi-\Omega t).
\end{equation}
The dumbbell branch is solved further assuming the separation of the variables $\cR(x,y) = \ell_\perp^{-2}(1- x^2/2) + \cR(y)$,
in which the equation reduces to
\begin{equation}
 \cR''(y) +\fr{2} \cR'(y)^2 + \cR(y) +\fr{2} \Omega^2 y^2 =0 \label{eq:dumbbell-master}
\end{equation}
where
\begin{equation}
 \ell_\perp^2 := \frac{2}{1+\sqrt{1-4\Omega^2}}.
\end{equation}
The dumbbells bifurcate from the onsets of the instability on the black bar at $\ell_\perp^{-2}\Omega^{-2}=2N$, where $N$ is a positive integer.
Since dumbbells produce $N$ bumps which develop to $N$ separate blobs, we refer to each branch as $N$-dumbbells.

\subsection{Blobs and necks}
\begin{figure}[t]
\begin{center}
\includegraphics[width=12cm]{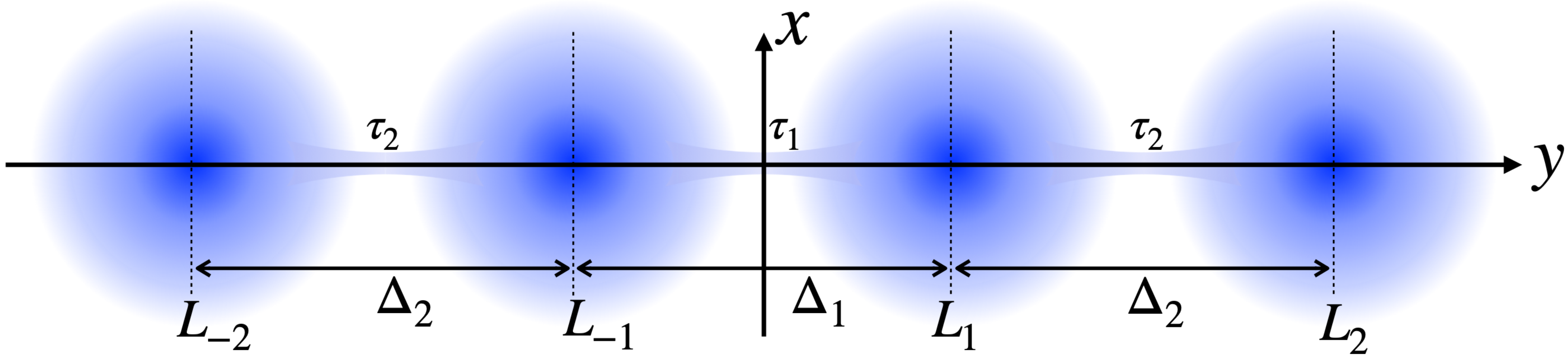}
\caption{Schematic picture of the mass density for $N=4$ dumbbell with the co-rotating coordinate. Thin neck structures exist between blobs.}
\label{fig:dumbscheme}
\end{center}
\end{figure}

In the limit of $\Omega \to 0$, eq.~(\ref{eq:dumbbell-master}) reduces to the black string effective equation~(\ref{eq:nubs-effeq}) and the solution is locally approximated by a Gaussian blob
\begin{equation}
 e^{\cR(y)} = e^{1-\fr{2}(y-L_i)^2} + \ord{\Omega^2}
\end{equation}
where $L_i$ is the peak position of each blob.
In the highly deformed phase, we assume a dumbbell consists of $N$ separate Gaussian blobs and the blobs are labeled
by $k= -k_{\rm max},\dots,k_{\rm max}:=\lfloor N/2 \rfloor$ from the negative end to the positive end (figure~\ref{fig:dumbscheme}). The index includes $k=0$ for odd $N$.

Now, we see the distance between blobs is determined by the balance between the neck tension
and rotation. 
By integrating eq.~(\ref{eq:dumbbell-master}) multiplied by $\cR'(z)e^{\cR(z)}$ from a neck waist $y_{\rm neck}$, we obtain
\begin{equation}
\left[ \left(\fr{2} \cR'(y)^2 + \cR(y)-1\right) e^{\cR(y)} \right]_{y_{\rm neck}}^{\infty} = -\frac{\Omega^2}{2}\int_{y_{\rm neck}}^{\infty} y^2e^{\cR(y)} \cR'(y) dy \label{eq:tension-dif}.
\end{equation}
Assuming the mass density falls off at $|y|\to \infty$, the right hand side reproduces the black string equation~(\ref{eq:nubs-effeq-tau}) near the neck, given the other part becomes constant. 
On the left hand side, the integral picks up the blob part and each blob integrals can be approximated by the full Gaussian integral, which gives the constant tension of $\ord{\Omega^2}$. 
As seen in the black string analysis, the effect of the tails will be the order of the tension which is now approximated to be $\ord{\Omega^2}$. Therefore, the cut-off tail part contributes to only $\ord{\Omega^4}$ in eq.~(\ref{eq:tension-dif}), and hence is safely negligible.

The approximated tension $\tau_k$ between $k-1$-th and $k$-th blob is given by
the sum of the integrals over Gaussian blobs at $y=L_{k},\dots,L_{k_{\rm max}}$,
\begin{equation}
- \left(\fr{2} \cR'^2 + \cR-1\right)  e^\cR  = \tau_k  =: \sqrt{2\pi} e \, \Omega^2\sum_{i=k}^{k_{\rm max}}  L_i .
\end{equation}
This can be interpreted as the balance condition between the brane tension and centrifugal forces on the outer blobs. By taking the difference, one can actually see the force balance on a single blob is given by the centrifugal force and tensions on both sides,
\begin{equation}
 \tau_{k} -\tau_{k+1} =   \cM_0 \, \Omega^2 L_k.
\end{equation}
where the Gaussian blob mass is given by $ \cM_0 =\sqrt{2\pi} e$ in eq.~(\ref{eq:nubs-gblob-mass}).\footnote{Here both the tension and Gaussian blob mass are one dimensional, since $x$-dependence is already factored out. }
We will expand this interpretation more precisely in section~\ref{sec:kinematic}.
By using the result of the black string analysis~(\ref{eq:tension-L-NLO}),
the local tension determines the interval $\Delta_k : = L_{k}-L_{k-1},$\footnote{In the higher order of $\Omega$,  eq.~(\ref{eq:tension-dif}) will give the coordinate dependence on the tension term due to the broken mirror symmetry,
in which the neck analysis should be changed from that of the black string.}
\begin{equation}
 \tau_k = \frac{ \Delta_k^2}{2} e^{1-\Delta_k^2/8}\left(1+\frac{2\pi^2}{3\Delta_k^2}\right),
\end{equation}
where we include the correction up to $\ord{\Delta_i^{-2}}$.
We note that $\Delta_1$ for even dumbbell is defined by $\Delta_1 = L_1-L_{-1}=2L_1$.
Eliminating the tension, we obtain a coupled equation for intervals,
\begin{equation}
\sqrt{2\pi} e \Omega^2\sum_{i=k}^{k_{\rm max}}  L_i =\frac{ \Delta_k^2}{2} e^{1-\frac{\Delta_k^2}{8}}\left(1+\frac{2\pi^2}{3\Delta_k^2}\right), \label{eq:dumbbell-cond}
\end{equation}
where
\begin{align}
& L_k =\left\{ \begin{array}{cc} \sum_{i=1}^k \Delta_i& (N: \ {\rm odd})\\ -\frac{\Delta_1}{2}+\sum_{i=1}^k \Delta_i&(N:\ {\rm even})\end{array}\right. .
\end{align}

\subsubsection{$2$-dumbbells}
Let us start from the easiest case, $N=2$.
Since we only have a single interval $\Delta_1 = 2L_1$, eq.~(\ref{eq:dumbbell-cond}) is no longer coupled,
\begin{equation}
  \sqrt{2\pi}  e \Omega^2 \frac{\Delta_1}{2} =  \frac{\Delta_1^2}{2}e^{1-\frac{\Delta_1^2}{8}}\left(1+\frac{2\pi^2}{3\Delta_1^2}\right).\label{eq:dumbbell-cond-2}
\end{equation} 
Ignoring the correction term in the left hand side, we obtain
\begin{equation}
 \Delta_1 = 2 \sqrt{-W_{-1}(-\pi\Omega^4/2)} \simeq 4 \sqrt{-\log \Omega}.\label{eq:2-dumb-delta-0}
\end{equation}
This reproduces the logarithmic dependence of $\Omega$ in the separation found in the numerical solution~\cite{Licht:2020odx}.
Including $\ord{\Delta_1^{-2}}$ term, we can improve the approximation,
\begin{equation}
 \Delta_1 = 2 \sqrt{-W_{-1}(-\pi\Omega^4/2)} \left(1+\frac{2\pi^2}{3W_{-1}(-\pi \Omega^4/2)^2}\right). \label{eq:2-dumb-delta-1}
\end{equation}

\subsubsection{Even dumbbells}
Now, we proceed to more general cases.
First, we consider even dumbbells with $N=2s$.
Since all the intervals are highly coupled in eq.~(\ref{eq:dumbbell-cond}), the equation is no longer directly solvable. 
In order to get hints for how to proceed, we have examined numerical solutions.
Following this, we assume the intervals are almost the same at $\Delta_k \to \infty$,
\begin{align}
 \Delta_k = \Delta +\ord{\Delta^{-1}}.
\end{align}
Then, each position is given by
\begin{equation}
 L_k \simeq \left( k-\fr{2} \right)\Delta.
\end{equation}
Eq.~(\ref{eq:dumbbell-cond}) now reduces to a single equation,
\begin{equation} 
 \sqrt{2\pi} \Omega^2 (s^2 -(k-1)^2)\Delta = \Delta^2 e^{-\frac{\Delta_k^2}{8}}.
\end{equation}
where $\Delta_k$ is kept in the exponent because the correction contributes to the factor.
This is solved by\footnote{Due to the logarithmic property, there is an ambiguity in the leading order solution that multiply $\Omega$ by a factor, which changes the next-to-leading order. One can also  include $k$-dependence in the next-to-leading order within the leading order solution as $\Delta_k = 2\sqrt{-W_{-1}[(s^2 -(k-1)^2)^2\Omega^4)/2]}$.
}
\begin{equation}
 \Delta_k = \Delta-\frac{4c_k}{\Delta} ,\qquad \Delta:=2 \sqrt{-W_{-1}(-\pi \Omega^4/2)},
\end{equation}
where
\begin{equation}
c_k=\log(s^2-(k-1)^2).
\end{equation}
Since $\Delta \sim \sqrt{-\log\Omega}$, this determines the blob position up to $\ord{1/\sqrt{-\log \Omega}}$.

For one higher order in $1/\Delta$, we need to take into account the contribution from the $k$-dependent correction. The blob positions are corrected by
\begin{equation}
 L_k = \left(k-\fr{2}\right)\Delta - \frac{4}{\Delta} \log\left( \frac{\Gamma(s+k)}{\Gamma(s-k+1)}\right).
\end{equation}
By the summation, we obtain
\begin{equation}
 \sum_{\ell=k}^{p} L_\ell = \fr{2}e^{c_k}\Delta- \frac{4}{\Delta} 
 \log \left(\frac{G(2s+1)}{G(s+k)G(s-k+2)}\right),
\end{equation}
where $G(n)$ is the Barnes G function defined by the superfactorial
\begin{equation}
 G(n) := \prod_{k=1}^{n-1} \Gamma(k).\label{eq:def-barnesG}
\end{equation}
Plugging this into eq.~(\ref{eq:dumbbell-cond}), we obtain $\ord{\Delta^{-3}}$ correction as
\begin{subequations}
\begin{equation}
 \Delta_k = \Delta - \frac{4c_k}{\Delta} + \frac{16d_k}{\Delta^3},
\end{equation}
where
\begin{align}
&c_k=\log(s^2-(k-1)^2),\\
& d_k = \frac{\pi^2}{6}- 2c_k- \fr{2}c_k^2 + 2 e^{-c_k} \log \left(\frac{G(2s+1)}{G(s+k)G(s-k+2)}\right).
\end{align}
\end{subequations}
 \begin{figure}[t]
\begin{center}
\includegraphics[width=12cm]{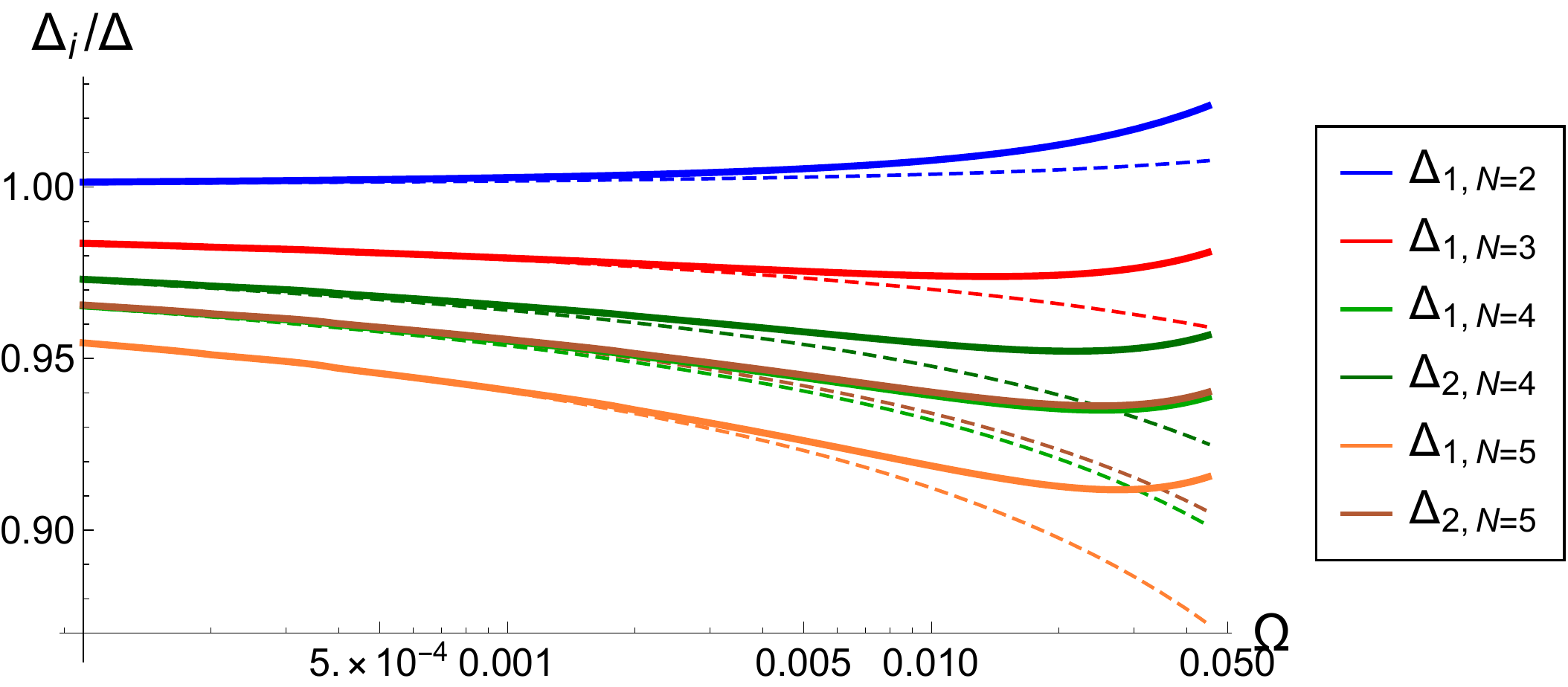}
\caption{The normalized separations are compared with the numerical result (solid curves). The dashed curves are the result up to $\ord{1/\Delta^3}$. The numerical data is produced by the same method used in ref.~\cite{Licht:2020odx}.}
\label{fig:dumbdelta}
\end{center}
\end{figure}

\subsubsection{Odd dumbbells}
Odd dumbbells with $N=2s+1$ can be similarly studied.
The interval is given by
\begin{subequations}
\begin{equation}
 \Delta_k = \Delta - \frac{4c_k}{\Delta} + \frac{16d_k}{\Delta^3},
\end{equation}
where $\Delta = 2 \sqrt{-W_{-1}(-\pi \Omega^4/2)}$ and
\begin{align}
& c_k = \log(s(s+1)-k(k-1)),\\
& d_k = \frac{\pi^2}{6}-  2c_k- \fr{2}c_k^2 + 2 e^{-c_k} \log \left(\frac{G(2s+2)}{G(s+k+1)G(s-k+2)}\right).
 \end{align}
 \end{subequations}
In figure~\ref{fig:dumbdelta}, we compare the result for several branches.

 \subsection{Phase diagram}
 Following the calculation in ref.~\cite{Licht:2020odx},
 the ratio of angular momentum to mass is given by
 \begin{equation}
  \frac{\cJ}{\cM} = \Omega \ell_\perp^2+\Omega\frac{\int_0^\infty y^2 e^{\cR(y)}dy}{\int_0^\infty e^{\cR(y)}dy}
 \end{equation}
Since the cut-off tails become sub-leading order in $\Omega$, the integrals are simply approximated by the sum of the Gaussian integrals over blobs at the leading order, which are easily evaluated for $N$-dumbbells as
 \begin{equation}
  \int_0^\infty e^{\cR(y)}dy \simeq \sqrt{2\pi} e N, \quad   \int_0^\infty y^2 e^{\cR(y)}dy \simeq \sqrt{2\pi} e \left(N+2\sum_{k=1}^{\lfloor N \rfloor} L_k^2\right).
 \end{equation}
 Using the expansion of $L_k$ by $\Delta = 2 \sqrt{-W_{-1}(-\pi\Omega^4/2)}$ up to $\ord{\Delta^{-1}}$, we obtain
 \begin{equation}
   \frac{\cJ}{\cM} =\frac{N^2-1}{12} \Omega  \Delta^2  \left( 1+\frac{j_1}{\Delta^2} +\ord{\Delta^{-4}}\right)+ \ord{\Omega^2},
 \end{equation}
and
 \begin{align}
j_1 = 1-\frac{192}{N(N^2-1)} \sum_{k=1}^{s} \left\{ \begin{array}{cl} (k-\fr{2})\log \frac{\Gamma(s+k)}{\Gamma(s-k+1)}& (N=2s)\\ k \log \frac{\Gamma(s+k+1)}{\Gamma(s-k+1)}& (N=2s+1) \end{array}\right. .
 \end{align}
  \begin{figure}[t]
\begin{center}
\includegraphics[width=10cm]{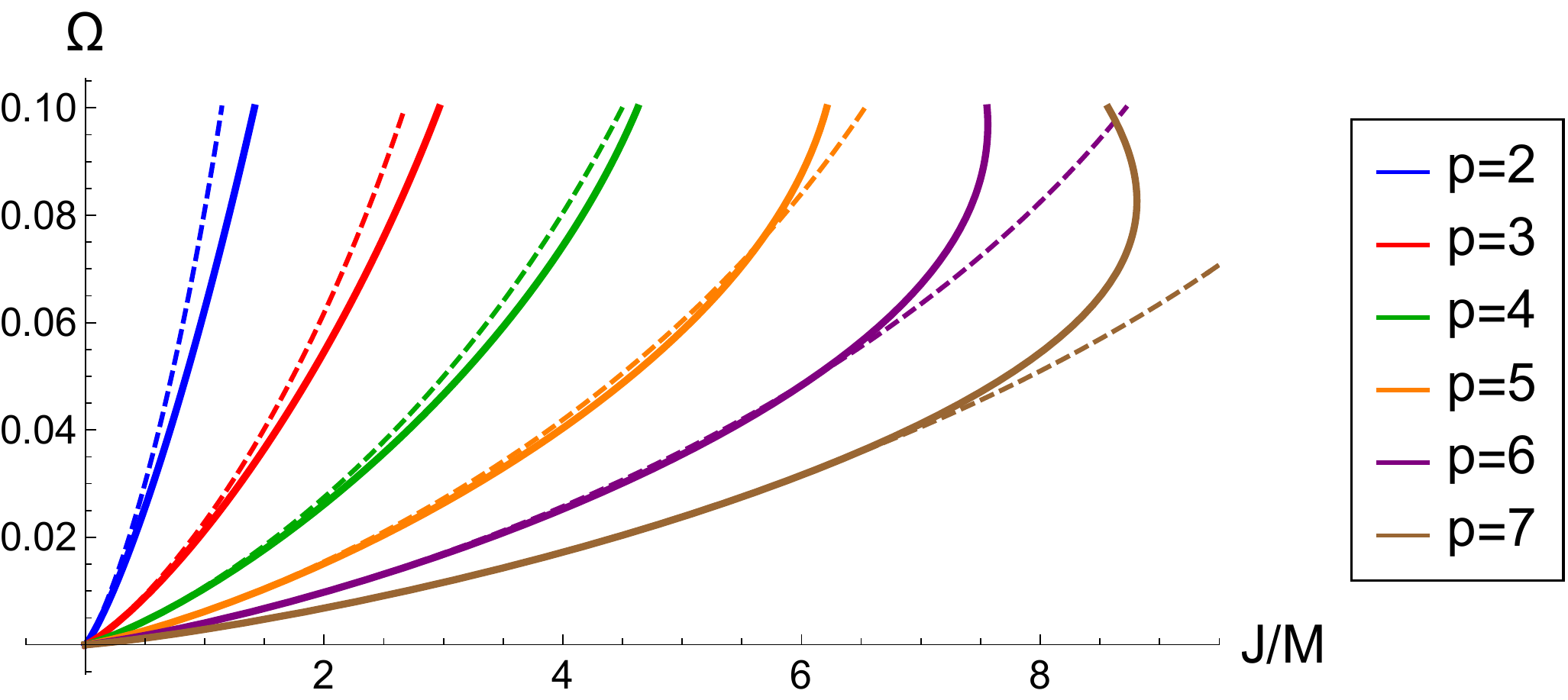}
\caption{The phase diagram of dumbbells. The approximation up to NLO in $1/\Delta$ (dashed curves) is compared with the numerical result (solid curves). The numerical data is produced by the same method used in ref.~\cite{Licht:2020odx}}
\label{fig:dumbJM}
\end{center}
\end{figure}
 Recalling the black string analysis, the blob approximation is only convergent for $\Delta \gtrsim 2\pi$, which is now corresponds to
 \begin{equation}
  \Omega = (2\pi)^{-\fr{4}} \Delta^\fr{2} \, e^{-\frac{\Delta^2}{16}} \lesssim 0.13.
 \end{equation}
 In figure~\ref{fig:dumbJM}, one can see the approximation reproduces the numerical result
 for $\Omega \lesssim 0.1$. Having more blobs worsens the approximation $L_{\rm max} \ll \Omega^{-1}$, and hence, dumbbells with larger $N$ have smaller convergent range.
 

\section{Highly deformed black ripples}\label{sec:ripple}
Black ripples, also called bumpy black holes, are axisymmetric deformed branches bifurcating from the zero modes of the ultraspinning instabilities on the Myers-Perry solution~\cite{Emparan:2003sy}. Numerical solutions have been actually constructed for $D=6,7$~\cite{Dias:2014cia,Emparan:2014pra}. In the large $D$ limit, black ripples are obtained by solving the $2+1$ axisymmetric, rigidly-rotating effective equation~\cite{Licht:2020odx},
\begin{equation}
 \cR''(r) + \fr{r} \cR'(r) + \fr{2} \cR'(r)^2 + \cR(r) + \fr{2} \Omega^2 r^2 = 0. \label{eq:ripple-master}
\end{equation}
Black ripples bifurcate from the onsets of the axisymmetric instability on the Myers-Perry solutions, which are given by 
\begin{equation}
 \Omega = \frac{\sqrt{2N-1}}{N}\quad \left( N = 2,3,4,\dots\right),
\end{equation}
where $N$ labels different branchs.
In ref.~\cite{Licht:2020odx}, a cluster of concentric Gaussian-shaped rings developed away from the axis in the highly deformed regime. This configuration, which we call {\it blob ring}, is another example of the blob solution. The number of blob rings is given by $s=\lfloor N/2 \rfloor$. Odd ripples also formed a central blob at the axis, which is isolated from blob rings.

In contrast to black strings and dumbbells, ripples involve two kinds of necks with separate length scales (figure~\ref{fig:rippleprofiles}). The axis region and innermost blob ring are connected by a thin long neck which scales as $\Omega^{-1}$ in the limit $\Omega \to 0$. 
The long neck takes a disk shape for even $N$ and ring shape for odd $N$, respectively.
Due to the dimensionality, these long neck region requires slightly different treatment.
On the other hand, the blob rings are connected with each other by (relatively) short necks which scales as $\sqrt{-\log\Omega}$ similar to the dumbbells.

\subsection{Blob rings}
From numerical solutions, we see that
even ripples are approximated by an array of several ring-shaped Gaussian profiles, {\it blob rings}, which are localized at $r \simeq \Omega^{-1}$ for $\Omega \to 0$. 
Blob rings themselves are also pulled apart from each other, with the distance logarithmically growing in $\Omega$ as in dumbbells.
Odd ripples have the same blob ring structure, but also form a central blob which is approximated by the Myers-Perry solution.

Now, we focus on one of the blob rings, which is located at $r=L \sim \Omega^{-1}$.
We take the limit $\Omega \to 0$, while keeping $w := \Omega L$ finite. By switching to the local coordinate $z:=r-L$,
the effective equation is expanded by $\Omega$,
\begin{equation}
  \cR''(z) + \fr{2} \cR'(z)^2 + \cR(z) + \frac{w^2}{2} = \sum_{i=1}^\infty \Omega^i \cS_i(z).
\end{equation}
 \begin{figure}[t]
\begin{center}
\includegraphics[width=15cm]{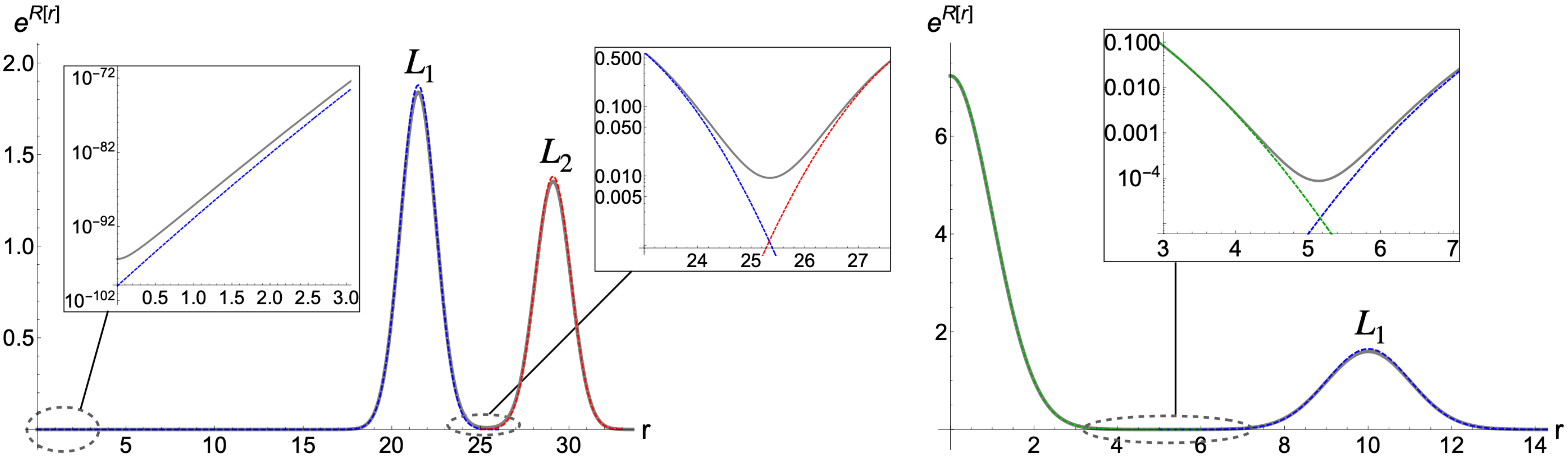}
\caption{The radial profile of ripples for $N=4$ with $\Omega = 0.04$ (left) and $N=3$ with $\Omega=0.1$ (right).
The blue and red curves are the Gaussian solution $\cR(r) = 1-\fr{2}\Omega^2 L_i^2-\fr{2}(r-L_i)^2$ with the fitted peak positions $L_i$. The green curve is the Myers-Perry solution.}
\label{fig:rippleprofiles}
\end{center}
\end{figure}

\paragraph{Leading order}
At the leading order, this reduces to the black string equation~(\ref{eq:nubs-effeq}) which admits the Gaussian solution
\begin{equation}
 \cR_0(z) = 1-\frac{w^2}{2}-\frac{z^2}{2}. \label{eq:ripple-LO-Gaussian}
\end{equation}
Given the value of $\Omega$ and the peak position $L$, this profile already gives a good fit to the numerical result for sufficiently small $\Omega$ ( figure~\ref{fig:rippleprofiles} ). 

Let us write the peak position of blob rings as $L_1<L_2<\dots<L_s$ and $w_i = \Omega L_i$, where $s$ is the number of the rings.
To determine the relation between $L_i$ and $\Omega$, we need an additional information about the global configuration. By integrating the master equation~(\ref{eq:ripple-master}), we obtain an useful equation
\begin{equation}
 \left[ \left(\fr{2} (\cR')^2 + R-1+\fr{2}\Omega^2 r^2\right) e^\cR \right]^\infty_r
 = \int^\infty_r \left(-\fr{r} (\cR')^2 + \Omega^2 r \right) e^\cR  dr. \label{eq:ripple-tension-integral}
\end{equation}
Assuming compactness, the left hand side should vanish for $r\to\infty$. 
On the other hand, inside the blob rings, the long tail from the innermost ring falls down until it meets the minimum on the very thin neck that is on the axis for even ripples, or halfway to the axis $r\simeq  \Omega^{-1}/2$ for odd ripples.
Therefore, if we integrate from the inner minimum to infinity, the left hand side is estimated as $\ord{e^{-\fr{2\Omega^{2}}}}$ for even ripples and $\ord{e^{-\fr{8\Omega^{2}}}}$ for odd ripples, both of which are non-perturbatively small  in $\Omega$.
On the right hand side, the integral over each single blob ring can be replaced by the full Gaussian integral
\begin{equation}
 \int_{i\sbar {\rm th}} \left(-\fr{r} (\cR')^2 + \Omega^2 r \right) e^\cR  dr \simeq \Omega \frac{\sqrt{2\pi} e^{1-\frac{w_i^2}{2}}}{w_i}(w_i^2-1), \label{eq:ripple-tension-integral-LO-1}
\end{equation}
where the cut-off tail turns out to be $\ord{\Omega^3}$. The integral from the inner minimum $r_{\rm min}$ to infinity is, then approximated by a sum of separate Gaussian integrals,
\begin{equation}
 \int_{r_{\rm min}}^\infty \left(-\fr{r} (\cR')^2 + \Omega^2 r \right) e^\cR  dr \simeq \Omega \sum_{i=1}^s \frac{\sqrt{2\pi} e^{1-\frac{{w_i}^2}{2}}}{w_i}(w_i^2-1).\label{eq:ripple-tension-integral-LO}
\end{equation}
We also assume that the blobs are close to each other.
\begin{equation}
 L_i-L_j \ll L_k \simeq \Omega^{-1}.
\end{equation}
Since eq.~(\ref{eq:ripple-tension-integral-LO}) should be non-perturbatively small, this sets $w_i \simeq 1$ and \footnote{If one allows $L_k - L_{k+1} =\ord{\Omega^{-1}}$, the left hand side again becomes non-perturbatively small $\ord{e^{-C\Omega^{-2}}}$ on the neck between $L_k$ and $L_{k+1}$ and then, one can to repeat the same analysis for smaller groups $\{L_1,\dots,L_k\}$ and $\{L_{k+1},\dots,L_s \}$, which leads to the same conclusion $L_i \simeq \Omega^{-1}$.}
\begin{equation}
 L_i \simeq \Omega^{-1} \left( 1+ \ord{\Omega}\right).
\end{equation}
This means that blob rings are sustained from collapse dominantly by centrifugal rotation, and the blob-blob interaction is secondary. It is worth noting that a similar hierarchical assumption was made for the study of multi-black rings/Saturns~\cite{Emparan:2010sx}.

\paragraph{Next-to-Leading order}
Now, we estimate the interaction between blob rings by solving the short necks between neighboring blob pairs.
Since the next-to-leading order source 
\begin{equation}
\cS_1(z) = - \fr{w}R'(z)+wz
\end{equation}
vanishes for the Gaussian solution~(\ref{eq:ripple-LO-Gaussian}) with $w_i=1+\ord{\Omega}$, the Gaussian solution remains the solution up to $\ord{\Omega}$. 
The leading order integral~(\ref{eq:ripple-tension-integral-LO-1}) is still valid up to $\ord{\Omega^2}$
\begin{equation}
 \int_{i\sbar {\rm th}} \left(-\fr{r}(R')^2+\Omega^2 r\right) e^R dr \simeq 2\sqrt{2\pi e} \,\Omega^2 \delta L_i,
\end{equation}
where we expanded by $w_i = 1+ \delta L_i\,\Omega$.
Then, eq.~(\ref{eq:ripple-tension-integral}) close to the neck between $i$-th and $i+1$-th blobs
reduces to the black string equation with the tension
\begin{equation}
 \tau_{i} = 2e\sqrt{2\pi}  \, \Omega^2  \sum_{j=i+1}^s \delta L_j+\ord{\Omega^3},
\label{eq:ripple-tension-rhs}
\end{equation}
where the scaling is adjusted by multiplying by $\sqrt{e}$.
Following the match between the neck waists and blob peaks in the black string analysis, the tension is given by a function of the distance between blobs~(\ref{eq:tension-L-NLO})
\begin{equation}
 \tau_{i} \simeq \frac{\Delta_{i}^2}{2} e^{1-\frac{\Delta_{i}^2}{8}}\left(1+\frac{2\pi^2}{3\Delta_{i}^2}\right)\label{eq:ripple-tension-lhs}
\end{equation}
where $\Delta_{i}:=\delta L_{i+1}-\delta L_i$ and terms up to $\ord{\Delta_i^{-2}}$ are taken. 
Evaluating the integral~(\ref{eq:ripple-tension-integral}) at the inner minimum, we obtain
\begin{equation}
 2\sqrt{2\pi e}\, \Omega^2 \sum_{i=1}^s \delta L_i \simeq (\cR_{\min}-1)e^{\cR_{\min}},
 \label{eq:ripple-dw-ave}
\end{equation}
where $\cR_{\rm min} \sim -\Omega^{-2}$ is the amplitude of the inner minimum.
As in the leading order analysis, the left hand side becomes non-perturbatively small in $\Omega$, and then, we obtain
\begin{equation}
\langle \delta L \rangle := \fr{s}\sum_{i=1}^s \delta L_i = \ord{\Omega,e^{-C\Omega^{-2}}},\label{eq:ripple-dw-ave-LO}
\end{equation}
where $\langle \cdot \rangle$ denotes the average over blob rings and $C$ is some constant.
In fact, eq.~(\ref{eq:ripple-dw-ave}) can be used to determine the non-perturbative corrections in $L_i$, which reflects the interaction between the long neck and blob rings. We will come back to this later.

Next, we study the short necks between blob rings. In the following, we assume $s>2$.
Equating eqs.~(\ref{eq:ripple-tension-rhs}) and (\ref{eq:ripple-tension-lhs}), 
a coupled equation about the separations between blobs,
\begin{equation}
2e\sqrt{2\pi}\,  \Omega^2  \sum_{j=i+1}^s \delta L_j
= \frac{\Delta_{i}^2}{2} e^{1-\frac{\Delta_{i}^2}{8}}\left(1+\frac{2\pi^2}{3\Delta_{i}^2}\right).
\label{eq:ripple-separations-con}
\end{equation}
This justifies the assumption that the cut off contribution becomes sub-leading order in $\Omega$.
As in the dumbbell analysis, we assume the separations are almost the same,
\begin{equation}
 \Delta_{i} \simeq \Delta+\ord{\Delta^{-1}}.
\end{equation}
Then, with the condition~(\ref{eq:ripple-dw-ave-LO}), we obtain
\begin{equation}
 \delta L_i = \frac{2i-s-1}{2}\Delta+\ord{\Delta^{-1}}.
\end{equation}
Plugging this into eq.~(\ref{eq:ripple-separations-con}), the leading order is determined
\begin{equation}
 \Delta = 2 \sqrt{-W_{-1}(-2\pi \Omega^4)}.
\end{equation}
Again, one can see the logarithmic growth in the separation $\Delta \simeq 4 \sqrt{- \log\Omega}$.
Expanding by $1/\Delta$, the sub-leading corrections are determined up to $\ord{\Delta^{-3}}$,
\begin{subequations}
\begin{equation}
 \Delta_i = \Delta - \frac{4c_i}{\Delta} + \frac{16d_i}{\Delta^3},
\end{equation}
where
\begin{align}
&c_i = \log(i(s-i)),\\
& d_i =\frac{\pi^2}{6} - 2 c_i - \frac{c_i^2}{2} + \frac{2}{i(s-i)} \log \left(\frac{G(s+1)}{G(i+1)G(s-i+1)}\right).
\end{align}
\end{subequations}
$G(n)$ is the Barnes G function defined in eq.~(\ref{eq:def-barnesG}).
By taking the sum and using $\langle \delta L \rangle=0$, the relative peak position is given by
\begin{equation}
 \delta L_i = \frac{2i-s-1}{2}\Delta- \frac{4}{\Delta} \log \left(\frac{\Gamma(i)}{\Gamma(s-i+1)}\right)
 +\frac{16}{\Delta^3} \left(\sum_{j=0}^{i-1} d_j\right),\label{eq:ripple-dL1-res}
\end{equation}
where
\begin{equation}
 d_0 := -\sum_{i=1}^{s-1} \frac{s-i}{s} d_i.
\end{equation}

\begin{figure}[t]
\begin{center}
\includegraphics[width=7cm]{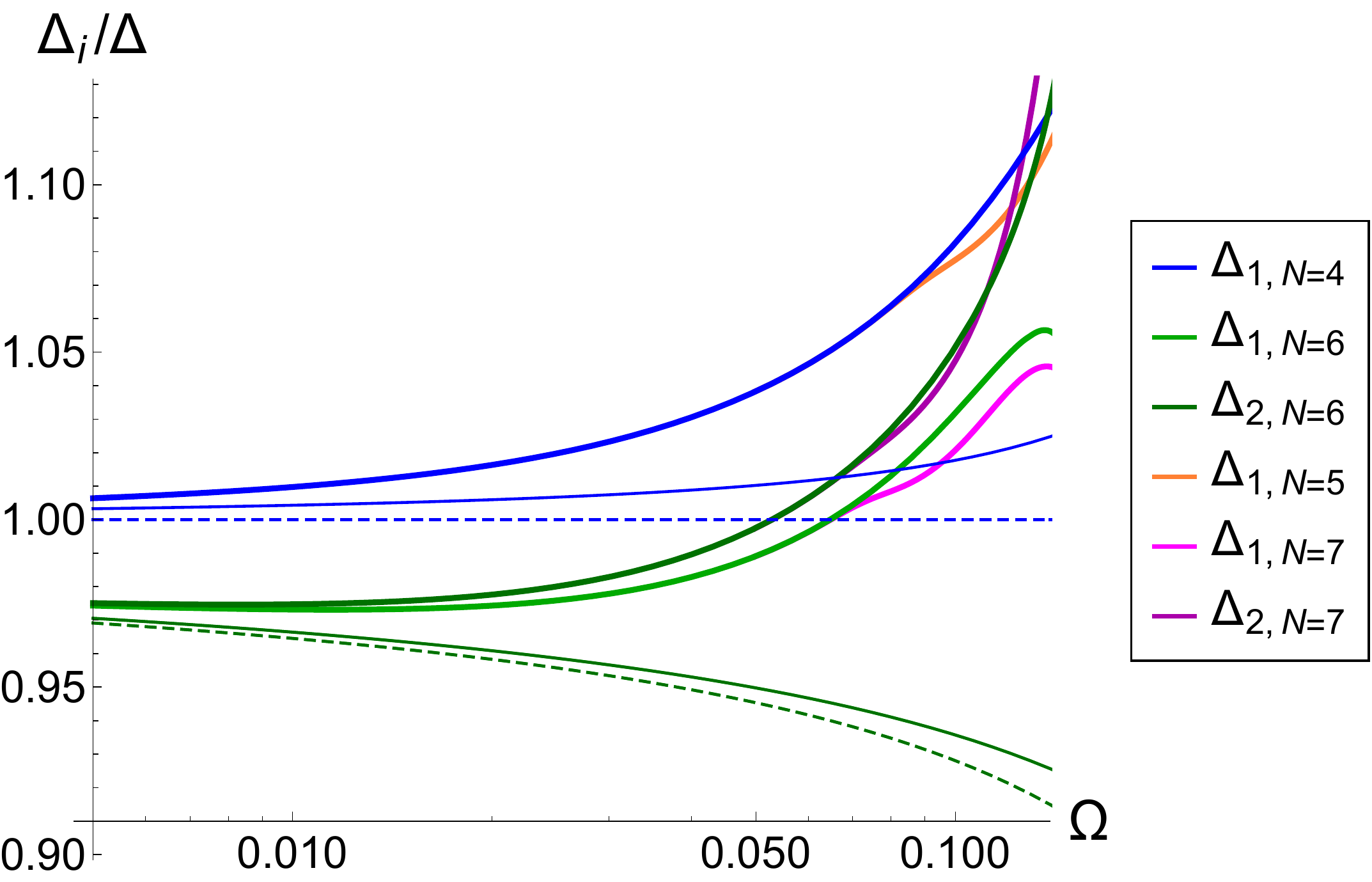}
\hspace{2mm}
\includegraphics[width=7.2cm]{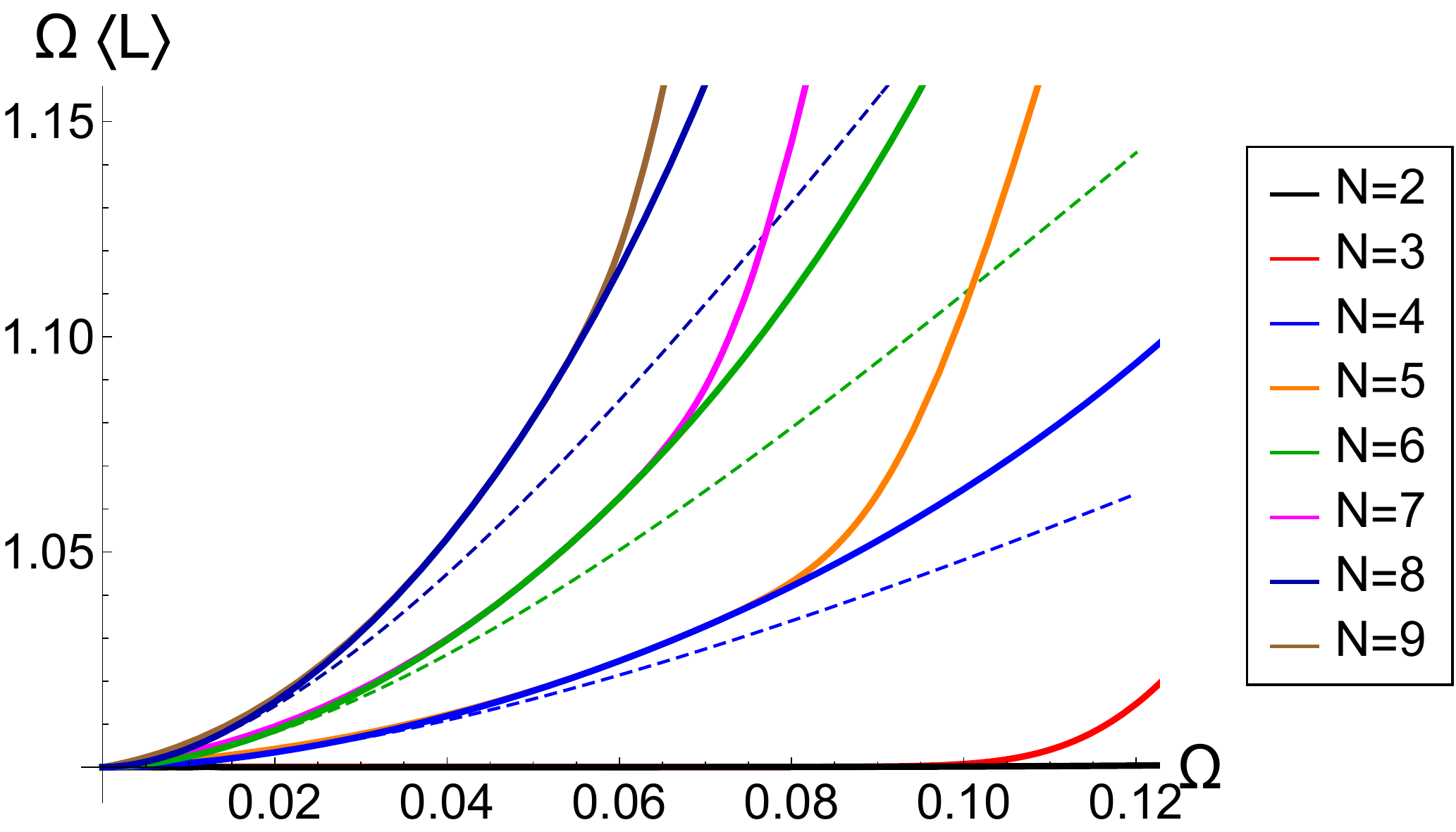}
\caption{The normalized separation between blob rings (left) and normalized average peak position (right). In each panel, the solid curve is the numerical result. 
In the left panel, NLO and NNLO results for $s=2,3$ are plotted as dashed and thin curves. The separations for $s=3$ are degenerate up to NNLO. In the right panel, the blob approximation up to $\ord{\Omega^2, \Delta^{0}}$ is shown as a dashed curve. The odd solutions share the same curve as the corresponding even ones until the point where non-perturbative effects become important. The numerical data are produced with the same method used in ref.~\cite{Licht:2020odx}.}
\label{fig:rippleL}
\end{center}
\end{figure}

\paragraph{Next-to-next-to-leading order}
One can continue determining the peak position to higher order in $\Omega$ by expanding further
\begin{equation}
 \delta L_i = \delta^{[1]}\! L_i + \delta^{[2]} \! L_i \,\Omega+ \dots.
\end{equation}
However, the calculation of $\delta^{[2]}\! L_i$ requires a detailed neck analysis with non-constant tension, to which the black string result no longer applies. Instead, we focus on the average of the correction which contributes to the mass and angular momentum,
\begin{equation}
\langle \delta^{[2]}\! L \rangle := \fr{s}\sum_{i=1}^s \delta^{[2]}\! L_i.
\end{equation}
By evaluating the integral~(\ref{eq:ripple-tension-integral}) up to $\ord{\Omega^3}$, we obtain
(see Appendix~\ref{app:ripple-dL2} for the detail),
\begin{equation}
\langle \delta^{[2]}\! L \rangle =  \fr{2} \langle \delta^{[1]} \! L^2 \rangle
 + \ord{\Delta^{-2}}, 
\label{eq:ripple-ave-dL2-res}
\end{equation}
where the squared average of $\delta^{[1]}\! L_i$ is calculated by using NLO result~(\ref{eq:ripple-dL1-res}),
\begin{align}
&\langle \delta^{[1]} \! L^2 \rangle := \fr{s}\sum_{i=1}^s (\delta^{[1]} \! L_i)^2 = \frac{s^2-1}{12} \Delta^2 - \frac{8}{s} \sum_{i=1}^s (2i-s-1) \log \Gamma(i)+\ord{\Delta^{-2}}.
\label{eq:ripple-ave-dLsq-res}
\end{align}
In figure~\ref{fig:rippleL}, the peak positions are compared with the numerical result for several branches.

\subsection{Long neck analysis}
\begin{figure}[t]
\begin{center}
\includegraphics[width=7.2cm]{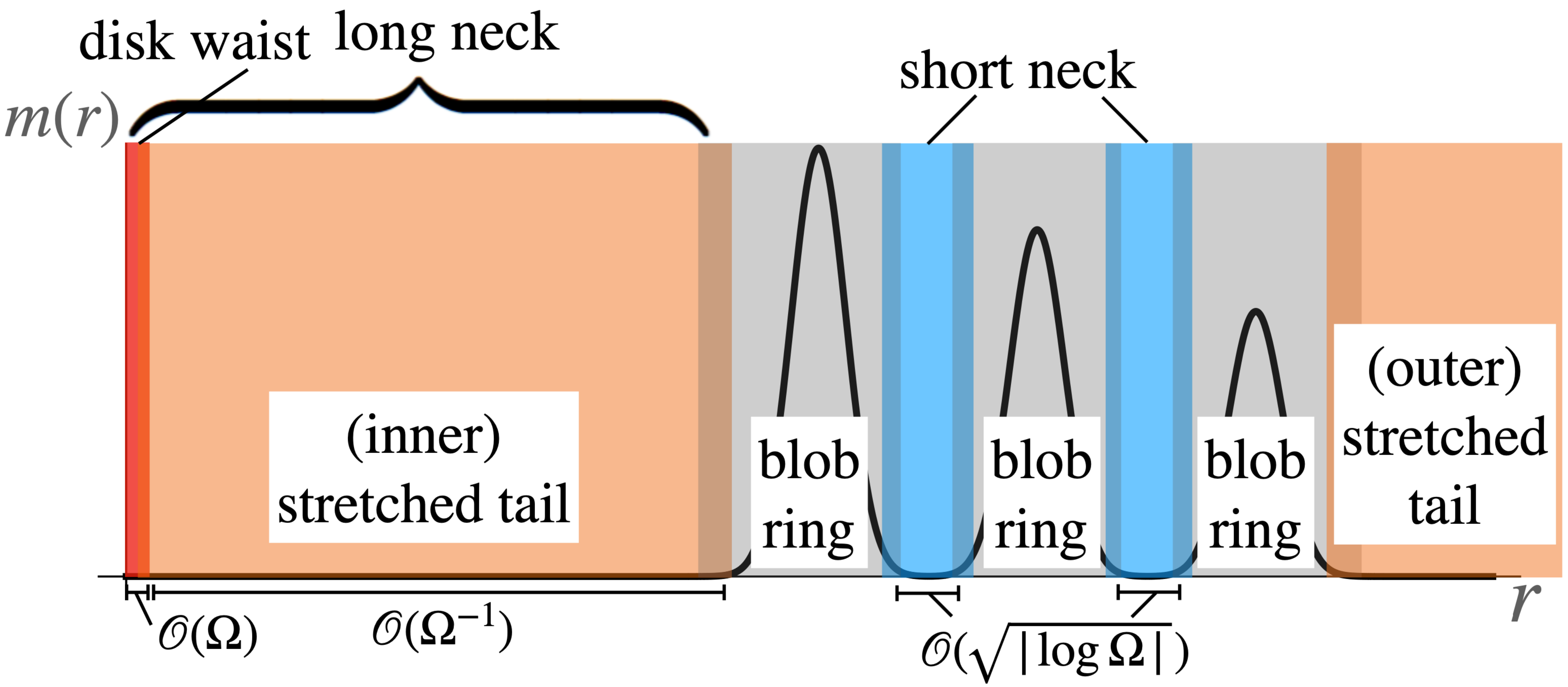}
\hspace{2mm}
\includegraphics[width=7.2cm]{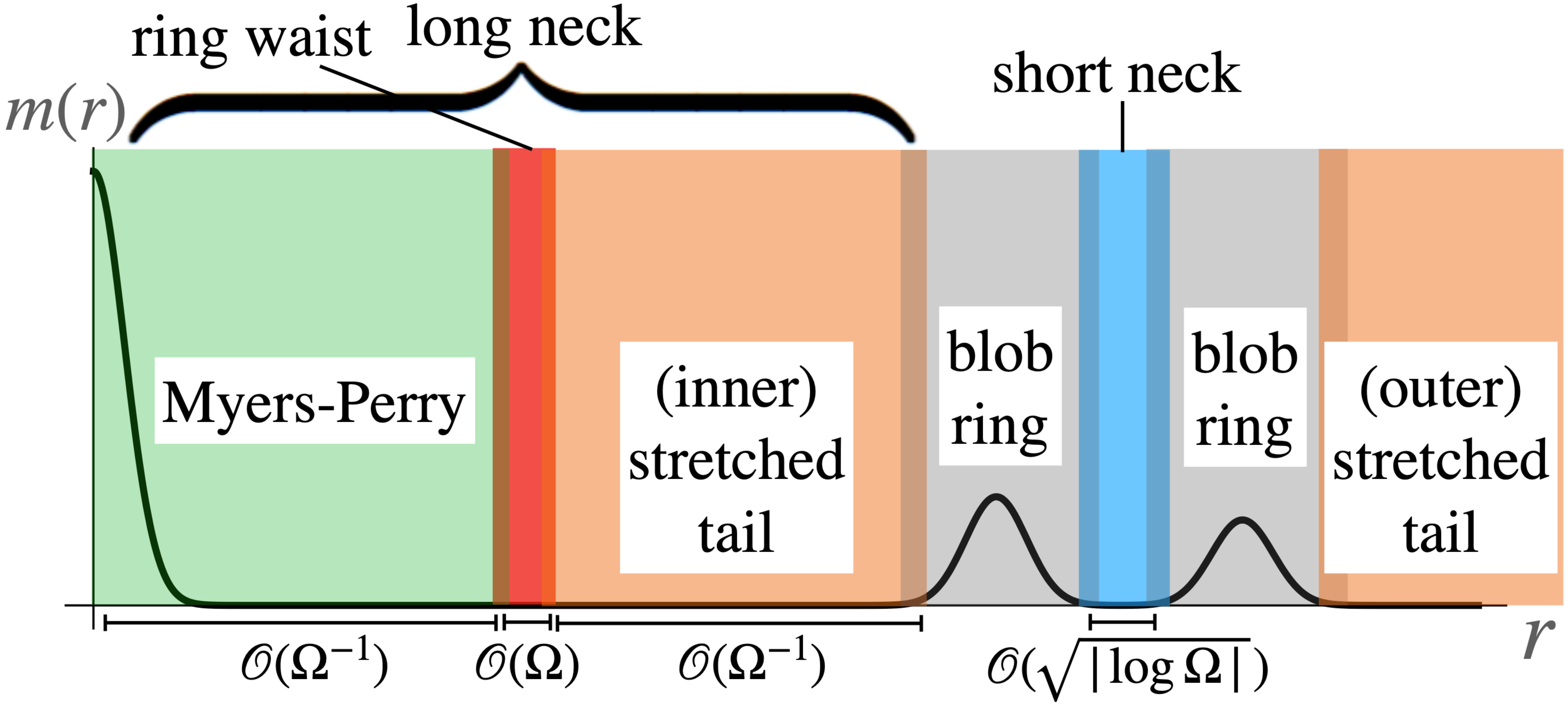}
\caption{Long neck structure and its scalings for even (left) and odd ripples (right).}
\label{fig:longneckpic}
\end{center}
\end{figure}

Up to now, the existence of the central blob did not matter to blob rings.
This is because the interaction decays exponentially in the distance of $\ord{\Omega^{-1}}$, 
only giving the magnitude of $e^{-C\Omega^{-2}}$ which is non-perturbative in $\Omega$.
This interaction can be captured by resolving the long neck and its waist using different patches (figure~\ref{fig:longneckpic}).

\subsubsection{Disk waist : even ripples}
First, we consider even ripples which consist of concentric blob rings without Myers-Perry blobs at the center.
Let us assume the peak of the innermost blob ring appears at $r = L$, which is $L \simeq \Omega^{-1}$.\footnote{We do not specify the detail form of $L=L(\Omega)$ throughout the long neck anaysis.}
Then, we observe a narrow disk-like waist close to the axis origin, whose magnitude is roughly estimated by the Gaussian tail as $\cR(r) \simeq -L^2/2 =\ord{\Omega^{-2}}$. As in the string neck, we zoom in on this region by rescaling
\begin{equation}
 \cR(r) = -\frac{L^2}{2} + \bar{R}(\xi),\quad \xi:= L r.
\end{equation}
The region $r = \ord{1}$ corresponds to $\xi \gg 1$ which allows the match at $\xi \to \infty$.
The effective equation is expanded by $\Omega^2$ with $w := \Omega L$ as
\begin{equation}
\bar{\cR}''(\xi)+\fr{2}\bar{\cR}'(\xi)^2+\fr{\xi}\bar{\cR}'(\xi)-\fr{2} =\sum_{i=1}^\infty \Omega^{2i} \bar{\cal S}_i
\end{equation}
from which we assume
\begin{equation}
 \bar{\cR}(\xi) = \sum_{i=0}\Omega^{2i} \bar{\cR}_i(\xi).
\end{equation}
One can see that the leading order equation now slightly differs from that of the black string due to the dimensionality.
Requiring the regularity on the axis, the leading order solution is given by the Bessel function
\begin{equation}
 \bar{\cR}_0(\xi) =\bar{r}_0 + 2 \log I_0(\xi/2),
\end{equation}
where $\bar{r}_0$ is the integration constant which sets the correction to the minimum on the axis,
\begin{equation}
 \cR_{\rm min} = - \frac{L^2}{2} + \bar{r}_0.\label{eq:ripple-axispinch-min}
\end{equation}
The leading order solution is expanded at $\xi\to\infty$ by
\begin{equation}
 \bar{\cR}_0(\xi) = \bar{r}_0 + \xi - \log \xi-\log \pi + \fr{2\xi}+\ord{\xi^{-2},e^{-\xi}}.
\end{equation}
The next-to-leading order equation is given by
\begin{equation}
\frac{d}{d\xi} \left(\xi I_0(\xi/2)^2 \frac{d}{d\xi} \bar{\cR}_1(\xi) \right) = - w^{-2}\xi I_0(\xi/2)^2\left(\bar{r}_0+2 \log I_0(\xi/2)\right).
\end{equation}
We cannot find the closed form solution at this order, but the asymptotic form is still available
\begin{equation}
 \bar{\cR}_1(\xi) =  - \frac{\xi^2}{2w^2} + \frac{\xi}{w^2}(-\bar{r}_0 +\log \pi + \log \xi)+O(\xi^0,e^{-\xi}).
\end{equation}
Combining the expansion up to NLO, the matching condition is obtained
by restoring the scaling while keeping $r = \ord{1}$,
\begin{subequations}
\begin{align}
& \cR \simeq-\frac{w^2}{2\Omega^2} + \bar{\cR}_0(\xi) + \Omega^2 \bar{\cR}_1(\xi)
\nonum
& \qquad = \bar{A}_0+\bar{A}_1 r -\frac{r^2}{2}+ \frac{2\Omega}{r} + \left(\Omega r-1\right) \log r +\ord{\Omega^2},
\end{align}
where
\begin{align}
&\bar{A}_0 = -\frac{w^2}{2\Omega^2}+\bar{r}_0 -\log \pi+\log \Omega, \\
&\bar{A}_1 = \frac{w}{\Omega} -( \bar{r}_0  + \log\Omega-\log \pi-\log w)\frac{\Omega}{w}.
\end{align}\label{eq:ripple-axistotail}
\end{subequations}
Therefore, the minimum on the axis~(\ref{eq:ripple-axispinch-min}) should be determined by matching to the outer region.

\subsubsection{Ring waist : odd ripples}
Odd ripples have a ring-type waist between the central blob and the innermost blob ring.
Assuming the minimum is located at $r = \mu \Omega^{-1}$, we consider the rescaling
\begin{equation}
 \cR(r) = -\frac{\mu^2}{2\Omega^2} + \bar{\cR}(\xi),\quad \xi:=\frac{\mu}{\Omega}(r-\mu\Omega^{-1}),
\end{equation}
where the solution is matched at $\xi\to \pm\infty$, which corresponds to the region $z:= r -\mu\Omega^{-1}=\ord{1}$.
The effective equation is expanded in $\Omega^2$ as
\begin{equation}
\bar{\cR}''(\xi)+\fr{2}\bar{\cR}'(\xi)^2-\fr{2} = \sum_{i=1}^\infty \Omega^{2i} \bar{\cal S}_i
\end{equation}
This is also solved by
\begin{equation}
\bar{\cR}(\xi) = \sum_{i=0}^\infty \Omega^{2i} \bar{\cR}_i(\xi).
\end{equation}
The solution up to NLO is given by
\begin{align}
&\bar{\cR}_0(\xi) =\bar{r}_0  + 2\log \cosh(\xi/2). \\
&\bar{\cR}_1(\xi) = -\frac{\xi}{\mu^2}+\frac{4}{\mu^2} \log \cosh(\xi/2)-\frac{2}{\mu^2} \text{Li}_2\left(-e^{-\xi}\right) \tanh (\xi/2)\nonum
&\quad +\fr{2\mu^2}\left(- \xi \left(2+\mu^2+2 \bar{r}_0  - 4\log 2\right)-\xi^2-\frac{\pi ^2}{3}+4\right)\tanh(\xi/2)
 \end{align}
 where $\bar{r}_0 $ gives the minimum at the waist.
 \begin{equation}
  \cR_{\rm min} = -\frac{\mu^2}{2\Omega^2} + \bar{r}_0 .\label{eq:ripple-ringpinch-min}
 \end{equation}
By fixing $z:= \Omega \xi/\mu$, the asymptotic expansion for $\xi\to \pm \infty$ leads to
\begin{subequations}
\begin{equation}
{\cal R}= \tilde{A}_{0} + \tilde{A}_{1,\pm} \,z -\frac{z^2}{2} + \ord{\Omega^2}
\end{equation}
where 
\begin{align}
 &\tilde{A}_{0} = - \frac{\mu ^2}{2\Omega^2}+\bar{r}_0  - 2\log 2,\\
 &\tilde{A}_{1,\pm} = \pm \frac{\mu }{\Omega} \pm \frac{\Omega}{2\mu} (2-\mu ^2-2\bar{r}_0 +4\log 2 \mp 2). 
\end{align}\label{eq:ripple-ringpinchtoeach}
\end{subequations}

\paragraph{Match with the central Myers-Perry blob}
The expansion at $\xi\to -\infty$ in eq.~(\ref{eq:ripple-ringpinchtoeach}) is easily matched with the Myers-Perry solution expanded by $ r = \mu \Omega^{-1} + z$,
\begin{equation}
{\cal R}_{\rm MP}(r) = \left(1+\sqrt{1-4\Omega^2}\right)\left(1-\frac{r^2}{4}\right) 
= -\frac{\mu^2}{2\Omega^2}+
2+\frac{\mu^2}{2}-\left(\frac{\mu}{\Omega}+\mu\Omega\right)z -\frac{z^2}{2}+\ord{\Omega^2},
\end{equation}
which determines
\begin{equation}
 \bar{r}_0  = \frac{\mu ^2}{2} + 2 + 2 \log 2+\ord{\Omega^2}.\label{eq:ripple-ringpinchtoMP-rm}
\end{equation}

\subsubsection{Stretched tail}
The matching region for the previous neck waists are still far from the blob rings at $r\simeq \Omega^{-1}$. In fact, we need another patch that feels the gradual change in the centrifugal barrier $\fr{2}\Omega^2 r^2$, since the coordinate patch for blob rings is too small for such global change.
It is worth noting that the center Myers-Perry blob in the odd ripple does not need such stretched patch, since it is a solution to the full effective equation, and hence already includes the rotational effect.
To this end, we impose a coarse-grained scaling to capture the Gaussian tail stretched by the rotation,
\begin{equation}
\cR(r) = \Omega^{-2} \tilde{\cR}(Y),\quad Y := \Omega r.
\end{equation}
The effective equation is rewritten as
\begin{equation}
\fr{2}\tilde{\cR}'(Y)^2+\tilde{\cR}(Y) =-\Omega^2 \left( \tilde{\cR}''(Y) + \fr{Y}\tilde{\cR}'(Y)+\frac{Y^2}{2} \right).
\end{equation}
The solution is obtained by the expansion in $\Omega^2$,
\begin{equation}
\tilde{\cR}(Y) = \sum_{i=1}^\infty \Omega^{2i} \tilde{\cR}_i(Y).
\end{equation}
The leading order solutions is solved by
\begin{align}
& \tilde{\cR}_0(Y) = - \fr{2}(Y-C)^2,
\end{align}
where $C$ is the integration constant. This constant sets the singular point in the subsequent equations. Here we set it at the first peak of the blob ring $C = w := \Omega L$ from which the tail appears. The next-to-leading order solution is obtained straightforwardly,
\begin{align}
 \tilde{\cR}_1(Y) = \frac{Y^2}{2} +\beta (Y-w) +\left(\frac{Y}{w}-1\right)\log \left(\frac{Y}{w}\right)
  + (1-w^2)\left(1+(w^2-1)\left(\frac{Y}{w}-1\right)\log\left|1-\frac{Y}{w}\right|\right)
\end{align}
where $\beta$ is an integration constant.

\paragraph{Match with the innermost blob ring}
First, we match the long tail to the innermost blob ring, which can be done regardless of whether $N$ is even or odd.
Switching to the coordinate from the peak $z:=r-w\Omega^{-1}=\Omega^{-1}(Y-w)$, the matching region with the innermost blob ring is given by $1\ll |z| \ll \Omega^{-1}$, and the matching condition becomes
\begin{align}
{\cal R} =1-\frac{w^2}{2} + \Omega (\beta+1) z - \fr{2} z^2 + \ord{\Omega^2}.
\label{eq:ripple-tailtoblob}
\end{align}
The match with eq.~(\ref{eq:ripple-LO-Gaussian}) fixes the translation degree of freedom,
\begin{equation}
 \beta = -1 + \ord{\Omega}.
\end{equation}
Note that one can also match with the outermost blob ring in the same manner.
We will not show this, since it does not have further information.

\paragraph{Match with disk waist}
Now, we consider the match on the axis for even ripples.
Assuming $r = Y/\Omega = \ord{1}$, we obtain
\begin{subequations}
\begin{equation}
  {\cal R} = \tilde{A}_0 + \tilde{A}_1 r -\frac{r^2}{2}  +\left( \Omega r-1\right)\log r + \frac{\Omega}{2 r}+\ord{\Omega^2}.
  \end{equation}
where
\begin{align}
 \tilde{A}_0 = -\frac{w^2}{2\Omega^2}-\beta-\log \Omega,\quad
\tilde{A}_1 = \frac{w}{\Omega}+(\beta+\log \Omega)\Omega.
\end{align}\label{eq:ripple-tailtoaxis}
\end{subequations}
The match between $\bar{A}_i$ in eq.~(\ref{eq:ripple-axistotail}) shows
\begin{align}
 \bar{r}_0 = \log \left(\frac{e\pi w}{\Omega^{2}}\right)+\ord{\Omega^2},
\end{align}
where we used $\beta \simeq -1$.
Therefore, the minimum on the axis for the even ripple is given by
\begin{equation}
\cR_{\rm even, \, min} = -\frac{w^2}{2\Omega^2} + \log \left(\frac{e\pi w}{\Omega^{2}}\right)+\ord{\Omega^2}.
\label{eq:ripple-even-min-res}
\end{equation}

\paragraph{Match with ring waist}
For odd ripples, 
the ring waist should be connected to both the long tail from the innermost ring as well as the center Myers-Perry.
Assuming the long neck minimum is at $r= \mu L = \mu \Omega^{-1}$, the matching condition is available by introducing
$z:= r-\mu \Omega^{-1} = \Omega ^{-1} (Y-\mu)$. Expanding by $\Omega$, we obtain
\begin{subequations}
\begin{align}
 {\cal R} = \tilde{A_0}+\tilde{A_1}  z- \fr{2} z^2 + \ord{\Omega^2},
\end{align}
where
\begin{align}
& \tilde{A}_0 = -\frac{(w-\mu)^2}{2\Omega^2}+\frac{\mu^2}{2}+\beta(\mu-w)\nonum
&\qquad+\left(\frac{\mu}{w}-1\right)\log \left(\frac{\mu}{w}\right)+(1-w^2)\left(1+\left(1-\frac{\mu}{w}\right)\log\left(1-\frac{\mu}{w}\right)\right),\\
&\tilde{A}_1 = \frac{w-\mu}{\Omega}+\left(\beta+\mu-\mu^{-1}+w+(w-w^{-1})\log \left(1-\frac{\mu}{w}\right) +w^{-1}\log\left(\frac{\mu}{w}\right) \right)\Omega.
\end{align}\label{eq:ripple-tailtoringpinch}
\end{subequations}
The match with $A_{i,+}$ in eq.~(\ref{eq:ripple-ringpinchtoeach}) determines
\begin{equation}
 \mu = \frac{w}{2} + \left(w+w^{-1}-\fr{2}w\log 2-\frac{1}{2}\right)\Omega^2+\ord{\Omega^4},
\end{equation}
and hence from eq.~(\ref{eq:ripple-ringpinchtoMP-rm}), we obtain
\begin{equation}
 \bar{r}_0 = \frac{w^2}{8} + 2 + 2 \log 2+\ord{\Omega^2}.
\end{equation}
Plugging these into eq.~(\ref{eq:ripple-ringpinch-min}), we obtain the minimum on the ring waist
\begin{equation}
\cR_{\rm odd, \, min} = -\frac{w^2}{8\Omega^2}+\frac{12+2w-3w^2}{8}+\frac{w^2+8}{4}\log 2 +\ord{\Omega^2}.
\label{eq:ripple-odd-min-res}
\end{equation}

\subsection{Phase diagram}
\begin{figure}[t]
\begin{center}
\includegraphics[width=13cm]{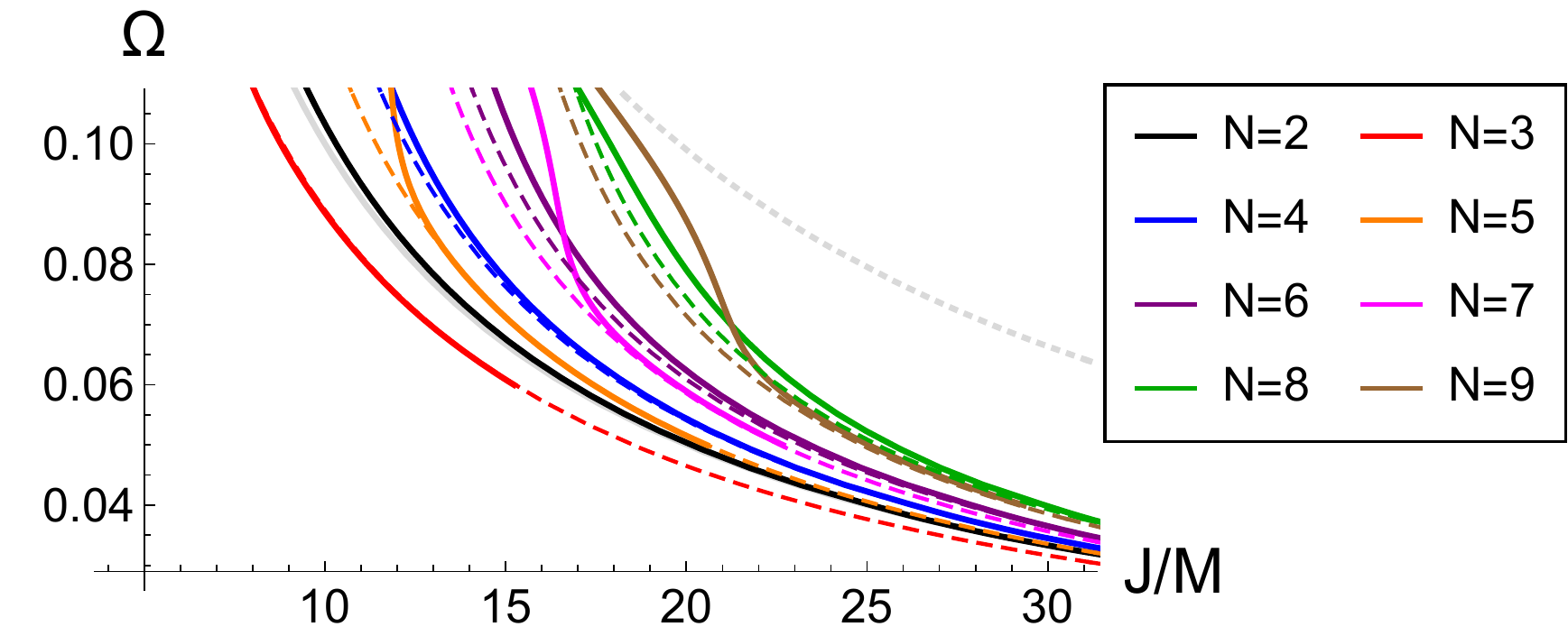}
\caption{Phase diagram of black ripples. The blob approximation (dashed) and the numerical result (solid) are compared. For reference, Myers-Perry black holes and Black bars are also plotted as gray dotted and solid curves. The numerical data are produced with the same method used in ref.~\cite{Licht:2020odx}.}
\label{fig:ripple-phase}
\end{center}
\end{figure}

For the ripples, the mass and angular momentum is given by
\begin{equation}
{\cal M} = 2\pi\int_0^\infty dr r e^{\cR(r)},\quad 
{\cal J} = 2\pi\Omega \int_0^\infty dr r^3 e^{\cR(r)}.
\end{equation}
Ignoring the non-perturbative terms, we can calculate the integrals up to $\ord{\Omega^2}$ by simply summing the contributions from all blobs
\begin{align}
&\frac{\Omega {\cal M}}{2\pi} = \sqrt{2\pi e} \, s \left(1 - \Omega^2  \langle \delta^{[1]} \! L^2 \rangle \right) + \veps_{\rm MP} e^2 \Omega - \sqrt{e}\sum_{i=1}^s \left( \int_{{\rm cut\, off}} e^{-\frac{z_i^2}{2}}dz_i -\Omega^2 \int \delta \cR^{(i)}(z_i) e^{-\frac{z_i^2}{2}} dz_i \right),\\
&\frac{\Omega^2 {\cal J}}{2\pi} =\sqrt{2\pi e} \, s\left( 1+ \Omega^2  \left(3+2\langle \delta^{[2]}\! L \rangle \right)\right) - \sqrt{e}\sum_{i=1}^s\left( \int_{{\rm cut\,off}} e^{-\frac{z_i^2}{2}}dz_i -\Omega^2 \int \delta \cR^{(i)}(z_i) e^{-\frac{z_i^2}{2}} dz_i \right).
\end{align}
where $\delta \cR^{(i)}$ is $\ord{\Omega^2}$ correction to the blob rings shown in eq.~(\ref{eq:app-ripple-dR}) and the cut off by the short necks between blobs also contributes to $\ord{\Omega^2}$.
However, since $\delta \cR^{(i)}$ and the cut off effects appear in the same manner in both quantities, they simply cancel out in the ratio.
$\veps_{\rm MP}$ corresponds to the existence of the central blob, which is set to be $1$ for the odd ripples and $0$ for the even ripples. 
Using eqs.~(\ref{eq:ripple-ave-dL2-res}) and (\ref{eq:ripple-ave-dLsq-res}), we obtain
\begin{align}
&\frac{\cal J}{\cal M} \simeq \fr{\Omega}+ \Omega \left( \frac{s^2-1}{6}\Delta^2 +j_1\right)
-\veps_{\rm MP}\left(\frac{e^\frac{3}{2}}{\sqrt{2\pi }\, s}-\frac{e^3 }{2\pi s^2}\Omega\right) ,\nonum
&\hspace{4cm} j_1 =3-\frac{16}{s} \sum_{i=1}^s (2i-s-1) \log \Gamma(i).
\label{eq:ripple-jm-res}
 \end{align}
In figure~\ref{fig:ripple-phase}, this is compared with the numerical results for several ripples.
One can see that odd ripples have less angular momentum per mass, because the central blob carries smaller angular momentum. With the larger number of blob rings, the central blob takes the smaller portion, and hence the reduction becomes smaller.

\subsection{Non-perturbative correction}
Here we evaluate the back reaction from the long neck tension which is non-perturbative in $\Omega$. Since, the short necks between the blob rings complicate the analysis, we consider
only $N=2$ and $N=3$ ripples which have only a single ring blob. 
Due to the absence of the short neck structure, we can easily solve only by expanding in $\Omega$ ( see appendix~\ref{sec:app-ripple-single} ). It turns out $w = 1 + \ord{\Omega^4,e^{-C\Omega^2}}$ and hence, eqs.~(\ref{eq:ripple-even-min-res}) and (\ref{eq:ripple-odd-min-res}) leads to
\begin{align}
& \cR_{{\rm min},N=2} =-\frac{1}{2\Omega^2}+\log\left(\frac{e\pi}{\Omega^2}\right),\\
& \cR_{{\rm min},N=3} =-\frac{1}{8\Omega^2}+\frac{11}{8}+\frac{9}{4} \log 2.
\end{align}
Substituting this to eq.~(\ref{eq:ripple-dw-ave}), we obtain the non-perturbative shift in the blob peaks
\begin{subequations}\label{eq:ripple-dL-NP}
\begin{equation}
 \delta L_{{\rm NP},N=2} = \fr{4}\sqrt{\frac{\pi}{2}}\,\Omega^{-6}  e^{\fr{2}-\frac{1}{2\Omega^2}},
\end{equation}
\begin{equation}
 \delta L_{{\rm NP},N=3} = \frac{1}{4 \times 2^{\frac{1}{4}}\sqrt{\pi}} \Omega^{-4} e^{\frac{7}{8}-\frac{1}{8\Omega^2}}.
\end{equation}
\end{subequations}

We also evaluate the non-perturbative correction to the phase diagram in $N=2$ and $N=3$ cases.
It turns out that both the cut off effect on the long neck and the non-perturbative correction to the blobs $\cR(r) = \cR_0(r)+\tau_{\rm NP} \delta \cR(r)$ contribute only in some positive power of $\Omega$ in the mass and angular momentum, which is much smaller than the non-perturbative change in the blob position~eq.~(\ref{eq:ripple-dL-NP}). Therefore, the leading non-perturbative correction is given by the non-perturbative shift in the blob peak,
\begin{equation}
 \frac{\cal J}{\cal M}  \simeq \left. \frac{\cal J}{\cal M} \right|_{{\rm pert}} + 2 \delta L_{{\rm NP}}.
\end{equation}
Using the single blob ring solution~(\ref{eq:app-ripple-single-ring}), the perturbative part is obtained up to $\ord{\Omega^{-9}}$,
\begin{align}
& \left. \frac{\cal J}{\cal M} \right|_{{\rm pert}, N=2,3} = \fr{\Omega}+3\Omega+9\Omega^3+63\Omega^5+617\Omega^7+\frac{22436}{3}\Omega^{9}\nonum
& -  \varepsilon_{\rm MP}\, \mu\left[1-\mu  \Omega+\left(\mu ^2+\frac{3}{2}\right) \Omega
   ^2 -\left(\mu ^3+2 \mu \right) \Omega ^3+\left(\mu ^4+\frac{5 \mu ^2}{2}+\frac{75}{8}\right) \Omega ^4- \left(\mu ^5+3 \mu ^3+12 \mu \right) \Omega ^5\right.\nonum
 &  \left.  \quad +\left(\mu ^6+\frac{7 \mu
   ^4}{2}+\frac{119 \mu ^2}{8}+\frac{3605}{48}\right) \Omega ^6- \left(\mu ^7+4 \mu ^5+18 \mu ^3+\frac{280 \mu }{3}\right) \Omega ^7 \right. \nonum
&\quad \left.+\left(\mu
   ^8+\frac{9 \mu ^6}{2}+\frac{171 \mu ^4}{8}+\frac{1821 \mu
   ^2}{16}+\frac{95091}{128}\right) \Omega ^8-\left(\mu ^9+5 \mu ^7+25 \mu ^5+\frac{410 \mu ^3}{3}+\frac{2690 \mu }{3}\right)
   \Omega ^9\right]\label{eq:ripple-n23-jmpert}
\end{align}
where $\veps_{\rm MP}$ takes $0$ for $N=2$ and $1$ for $N=3$.
For the compactness, we write $\mu := e^{3/2}/\sqrt{2\pi}$.
%
\begin{figure}[t]
\begin{center}
\includegraphics[width=7.2cm]{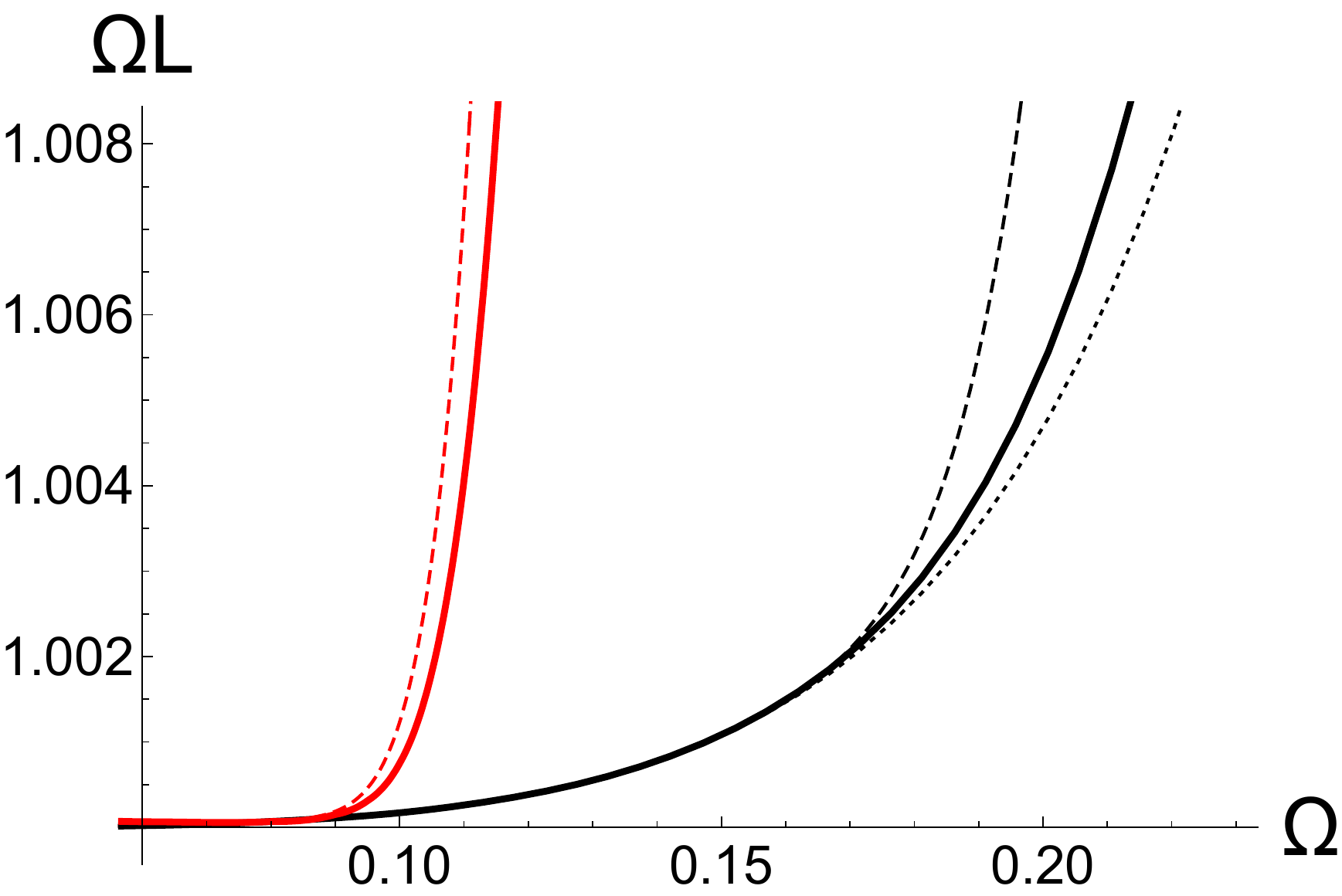}
\includegraphics[width=7.2cm]{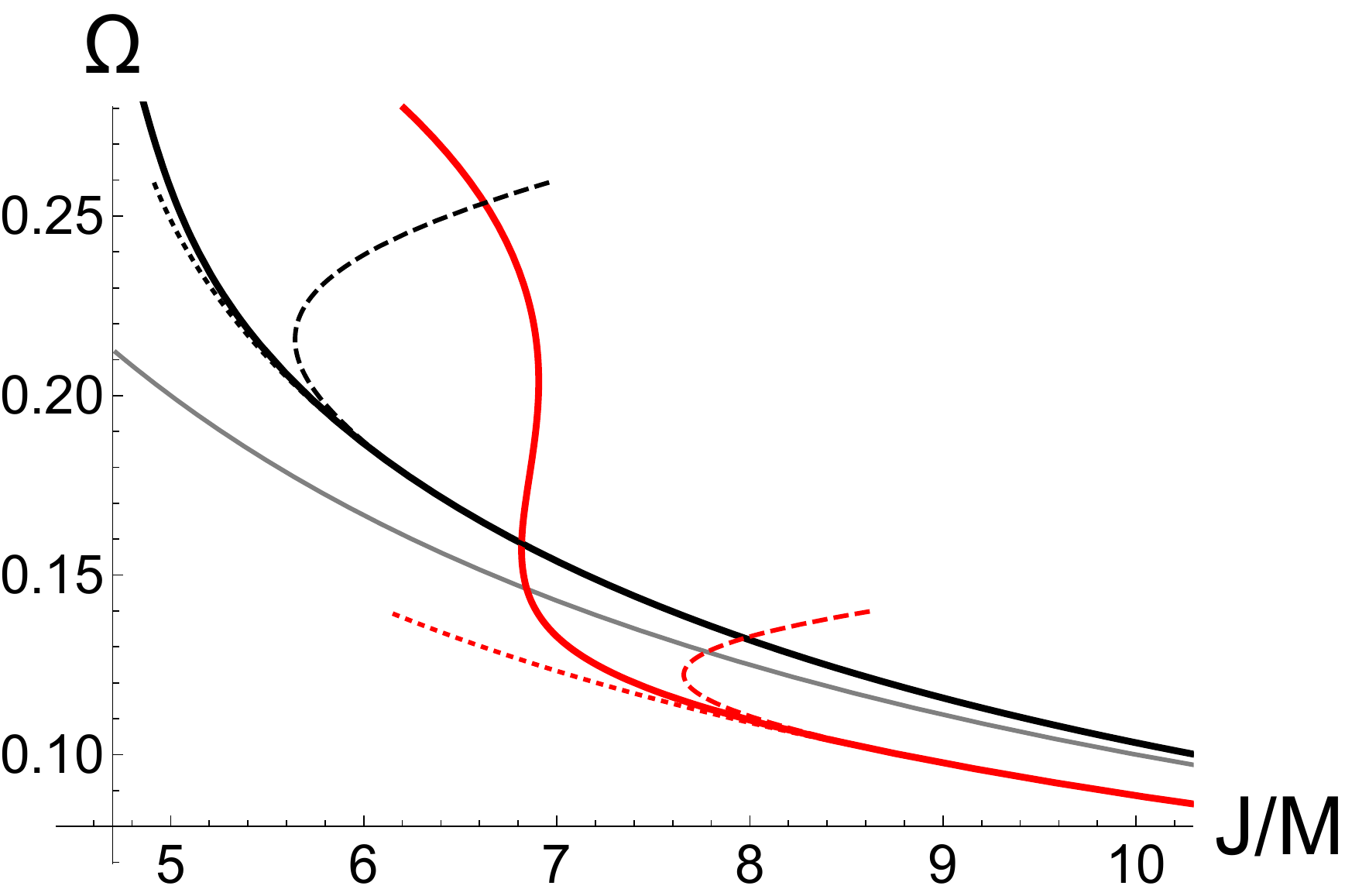}
\caption{The normalized peak positions of blob ring and phase diagram for $N=2$ and $N=3$. The blue and red solid curves are the numerical result for $N=2$ and $N=3$. The dashed and dotted curves with corresponding colors are the blob approximation with and without non-perturbative shift. The perturbative part is expanded up to $\ord{\Omega^{10}}$ as in eq.~(\ref{eq:app-ripple-single-L-pert}), so that the expansion is convergent for $\Omega \lesssim 0.2$. In the right panel, the black bars are also plotted by the gray curve for the reference.}
\label{fig:ripple-L23NP}
\end{center}
\end{figure}

In figure \ref{fig:ripple-L23NP}, one can see that the rise of the non-perturbative correction corresponds to the sudden inflection around $\Omega \sim 0.18$ for $N=2$ and $\Omega \sim 0.1$ for $N=3$ both in the peak position and phase diagram.
Unfortunately, since the factor of the non-perturbative correction is determined only up to the leading order in $\Omega$, the curves after the inflection do not fit well the numerical data.
In the phase diagram, the non-perturbative effects lead to an increase in ${\cal J}/{\cal M}$.
This is because the attraction between blobs requires larger separation to balance it.

\subsection{Blob rings and black rings}
Lastly, the single blob ring result is compared with the other black ring results.
The comparison indicates that in a certain range of $\Omega$
blob rings describe black rings/Saturns in the same manner as Myers-Perry blobs do for Myers-Perry black holes.

\paragraph{Blackfold at large $D$}
It turns out that the perturbative correction in the blob ring~(\ref{eq:ripple-n23-jmpert}) corresponds to  the finite size effect from the elastic bending energy for the black ring solved in the blackfold approach.
For $n\to\infty$, eq.(32) in ref.~\cite{Armas:2014bia} becomes
\begin{equation}
 \omega_H = \fr{2j} \left(1+\frac{3}{4nj^2}\right)\label{eq:ripple-fromblackfold},
\end{equation}
where the normalized angular velocity and angular momentum at the large $D$ are rewritten by the current variables as
\begin{equation}
 \omega_H \to \frac{n}{4\pi T } \sqrt{n} \Omega,\quad j \to \frac{4 \pi T}{n} \frac{\cJ}{2\sqrt{n}\cM},
\end{equation}
where we also used ${\cal J}/{\cal M}|_{\rm phys} = {\cal J}/{\cal M}/n$ due to the spacial rescaling by $1/\sqrt{n}$ in the brane setup~(\ref{eq:eft-ansatz}).
Recalling that the large $D$ effective theory has $T=n/(4\pi)$ at the leading order, eq.~(\ref{eq:ripple-fromblackfold}) coincides with the terms up to $\ord{\Omega}$ in eq.~(\ref{eq:ripple-n23-jmpert}).
This coincidence would imply the equivalence between the blob approximation and blackfold at large $D$.
It is also interesting to see whether the non-perturbative effect coincides with the self-gravitational effect in the blackfold approach.

\paragraph{Slowly rotating limit}
In the slowly rotating limit $\Omega = \ords{1/\sqrt{D}}$, the $1/D$ expansion around the long neck waist will break down, 
indicating that the effective theory needs to be identified with another coordinate patch.
Assuming the slowly rotating ansatz, the large $D$ black rings were also solved in the effective theory on the ring coordinate~\cite{Tanabe:2015hda,Chen:2017wpf,Chen:2018vbv}\footnote{AdS background also allows fast rotating rings with $\Omega = \ord{D^0}$~\cite{Mandlik:2020lgf}.}.
Here, instead of the detailed match, we simply present an evidence that the blob ring in the slow rotating limit can be identified as a part of the ring coordinate. First, we evaluate the physical quantities of the asymptotically flat, vacuum large $D$ black ring~\cite{Tanabe:2015hda} at the leading order, which were written in the integral form,
\begin{align}
& {\cal M} = \frac{n \Omega_n}{8G_D} \int_{-1}^1 dy \frac{(R^2-1)e^{P(y)}}{R\sqrt{1-y^2}} \left(\frac{R\sqrt{1-y^2}}{R+y}\right)^n,\\
&  {\cal J} = \frac{\sqrt{n} \Omega_n}{8G_D} \int_{-1}^1 dy \frac{(R^2-1)^{5/2} e^{P(y)}}{R(R+y)^2\sqrt{1-y^2}} \left(\frac{R\sqrt{1-y^2}}{R+y}\right)^n,\\
& \Omega_H = \fr{\sqrt{n}} \frac{\sqrt{R^2-1}}{R^2}, \label{eq:ripple-tanabe-omega}
\end{align}
where $P(y)$ is a regular function.
 One should note that $R$ here is a parameter related to the ring radius but not an observable, and hence should be eliminated in the comparison.
To extract the leading order contribution from the integrals, we expand the integrand around the extremum in the $n$-powered term,
\begin{equation}
 y = -\fr{R} + \frac{R^2-1}{R^2}\frac{z}{\sqrt{n}}. \label{eq:ripple-tanabe-rescale}
\end{equation}
The limit $n\to\infty$ leads to
\begin{align}
&\cM \simeq \cM_0 \int_{-\infty}^\infty dz e^{-\frac{z^2}{2}} = \sqrt{2\pi} \alpha, \\
&\cJ \simeq \frac{\alpha R^2}{\sqrt{n}\sqrt{R^2-1}} \int_{-\infty}^\infty dz e^{-\frac{z^2}{2}}
 = \frac{\sqrt{2\pi}\alpha}{n\, \Omega},
\end{align}
where eq.~(\ref{eq:ripple-tanabe-omega}) is used, and the following constant is fixed in the limit
\begin{equation}
 \alpha := \frac{\sqrt{n} \Omega_n}{8G_D} \left(\frac{R}{\sqrt{R^2-1}}\right)^{n-2} 
  \sqrt{R^2-1}\, e^{P(-1/R)}.
\end{equation}
Taking into account the difference in the scaling by $1/n$ in the definition of ${\cal J}$, this coincides with the phase diagram of the single blob ring up to the leading order.\footnote{The same relation is obtained by using the large $D$ charged ring in ref.~\cite{Chen:2017wpf}.}
Besides, one can notice that eq.~(\ref{eq:ripple-tanabe-rescale}) locally rescales the coordinate that of the brane effective theory, instead of the compact polar coordinate.
Therefore, the small region around $y=-1/R$ should correspond to the slowly rotating blob ring, since the leading contribution comes from there.


\section{Kinematic motion of blobs}\label{sec:kinematic}
In ref.~\cite{Andrade:2019edf,Andrade:2018yqu,Andrade:2020ilm}, 
it was observed that during a collision process blobs are attracted each other.
In some cases, they even orbit several rounds before the collision.
Here we show these resurgent kinematic features in the effective theory
are described by the neck interaction.

Let us consider several traveling blobs sufficiently separate from each other in the $p+1$-effective thoery~(\ref{eq:eft-effectiveLO}).
Then, we divide the space domain into several regions, so that each region covers each blob.
The boundaries run on the pits between them, which will pass through the neck waists.
For simplicity, we assume the time dependence of the boundaries is absent or negligible.
By integrating the quasi-local stress tensor~(\ref{eq:eft-stresstensor}) over each region, the physical quantities of each blob are defined,
\begin{align}
& {\cal M}_I(t) := \int_{V_I} T^{tt} dV,\quad {\cal P}_I^i(t) := \int_{V_I} \, T^{ti} dV,
\end{align}
where $I=A,B,C,\dots$ denotes each blob and corresponding region. Obviously, the total sum of each quantity recovers total conserved quantities.
We also define the blob position by
\begin{equation}
 X_I^i(t) := \fr{{\cal M}_I(t)}\int_{V_I} x^i\, T^{tt} dV,
\end{equation}
which strictly gives the peak position for a single Gaussian blob.
By using the conservation equation with fixed boundary, the change in each quantity is expressed by
\begin{align}
& \dot {\cal M}_I(t) = -  \int_{\partial V_I} m v^i dS_i,\nonum
& \dot X_I^i(t) = {\cal P}_I^i(t) - \fr{{\cal M}_I(t)} \int_{\partial V_I} \left(x^i-X_I^i(t)\right)mv^j dS_j\nonum
& \dot {\cal P}_I^i(t) = - \int_{\partial V_I} (m v^i v^j +\tau^{ij}) dS^j,
\end{align}
where the dot denotes the time derivative.
Assuming that the boundary flux terms are sub-leading order, each blob mass 
becomes constant, and the forces on blobs only come from the boundary stress tensor 
\begin{align}
&{\cal M}_I \ddot X_I^i(t) =- \int_{ \partial V_I} 
\tau^{ij} dS_j.\label{eq:kin-tensionforce}
\end{align}
\begin{figure}[t]
\begin{center}
\includegraphics[width=8cm]{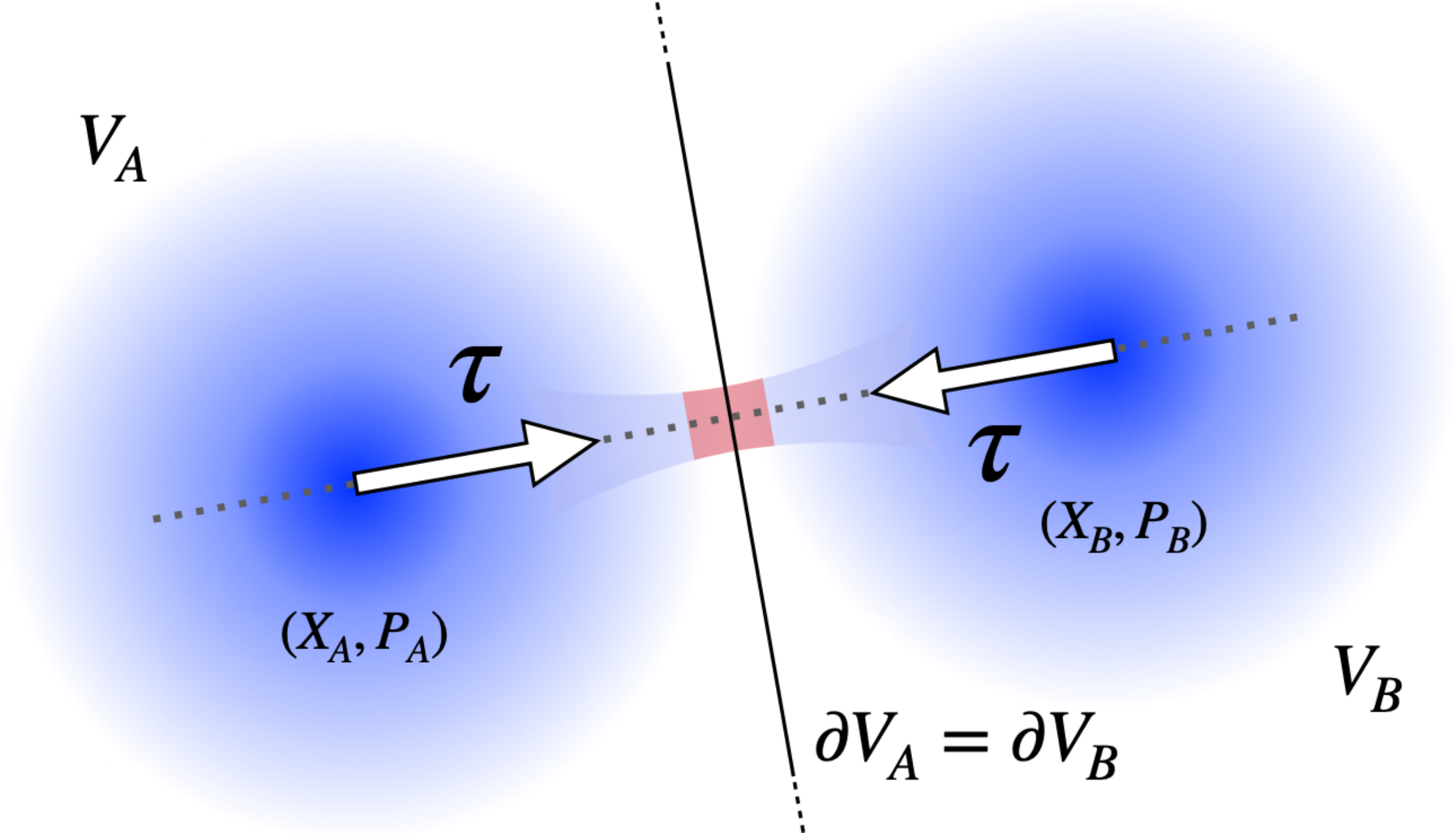}
\caption{Interaction between two equal mass blobs. The dominant contribution of the boundary integral comes from the tension on the neck waist (red region), whose cross section (dotted line) is approximated by that of the black string neck. }
\label{fig:blobinteraction}
\end{center}
\end{figure}

Now, we consider the interaction between two blobs $A$ and $B$ (Figure~\ref{fig:blobinteraction}). 
For simplicity, we assume the two have the equal mass.
Since the blobs are separate from each other, on the straight line from one peak to another peak,
 the solution looks like that of a black string. From the symmetry around that line, it is natural to expect the solution still presents Gaussian profiles in the other directions,
  \begin{equation}
  m \simeq e^{\cR(z)} e^{-\frac{|{\bf y}|^2}{2}}, \label{eq:kin-twoblobs-m}
 \end{equation}
 where $z$ is the coordinate along the peak-to-peak line and ${\bf y}$ other perpendicular directions. We assume $\cR(z)$ follows eq.~(\ref{eq:nubs-effeq}) with the appropriate mass scale $C$.
 Similarly, the neck tension on the waist would also be approximated by
\begin{equation}
 \tau_{zz} \simeq -e^C \tau \, e^{-\frac{|{\bf y}|^2}{2}},\quad \tau_{ij} \simeq 0  {\rm \ for \ other} \ (i,j) ,
\end{equation}
where $\tau$ is given by eq.~(\ref{eq:tension-L-NLO}) and $C$ is the mass scale. 
Since both blobs are closely approximated by Gaussian blobs, the masses are given by that of the Gaussian blob mass on the $p$-brane,
\begin{equation}
 {\cal M}_A = {\cal M}_B \simeq (2\pi)^\frac{p-1}{2} \int_{-\infty}^{\infty} dz e^{1+C-\frac{z^2}{2}} = (2\pi)^\frac{p}{2} e^{1+C}=:  \cM_0 
\end{equation}
Assuming the blob $A$ is in the positive side and $B$ in the negative side of $z$, 
eq.~(\ref{eq:kin-tensionforce}) gives
the equation of motion for blobs
\begin{equation}
 \ddot{X}^z_A(t) = -  \ddot{X}^z_B(t)  
\simeq -\frac{(2\pi)^\frac{p-1}{2} e^C \tau}{ \cM_0 }  \simeq - \frac{L_{AB}^2}{2\sqrt{2\pi}} e^{-\frac{L_{AB}^2}{8}},
\end{equation}
where $L_{AB}=|X_A-X_B|$ and we just take the leading order term in eq.~(\ref{eq:tension-L-NLO}) for simplicity. Since this force goes from a peak to another peak, we can rewrite this in more general form
\begin{equation}
 \cM_0 \, \ddot {X}_I  = - \sum_{J\neq I}F(|X_I-X_J|)\frac{X_I-X_J }{|X_I-X_J|},
 \label{eq:kin-force-I}
\end{equation}
where the index $J$ runs for all other neighboring blobs and the attraction force is given by
\begin{equation}
 F(r) = \frac{ \cM_0 \, r^2}{2\sqrt{2\pi}} e^{-\frac{r^2}{8}}.\label{eq:kin-equal-centralf}
\end{equation}
This force seems to give a nonlinear elastic force with the yield point at $r=2\sqrt{2}$.
However, one should recall that the formula~(\ref{eq:tension-L-NLO}) does not converge
already below $r \sim 2\pi$. Instead, one can expect the blobs start merging inside that radius.
In~\cite{Andrade:2020ilm}, the outgoing impact parameter after the fission is estimated as $b_{\rm out} \approx 8$.
This corresponds to the fact that the tension decays to almost zero near $r \simeq 8$.
The fussionless $2\to2$ events are also observed for some part of $b_{\rm in} \gtrsim 4$. 
However, there is no clear distinction between $2\to1\to2$ and $2\to2$. These fusionless events
could also be due to the exponentially decaying tension.

If the two blobs are stationary, rotating with angular velocity, then the tension should
balance with the centrifugal force. For the stationary motion, it is convenient to introduce the center mass frame, in which the coordinates become
\begin{equation}
 X_A(t) = {\bf r}(t)/2,\quad X_B(t) = -{\bf r}(t)/2,
\end{equation}
and the equation of motion is rewritten as
\begin{equation}
  \fr{2} \cM_0  \, \ddot {\bf r}(t) =  -F(r) \hat{\bf r}, \quad (r:=|{\bf r}|,\quad \hat{\bf r} := {\bf r}/|{\bf r}|).
\end{equation}
The stationary balance condition becomes
\begin{equation}
   \cM_0 \, \Omega^2 \frac{r}{2} = F(r).
\end{equation}
This actually reproduces the balance condition for $2$-dumbbells~(\ref{eq:dumbbell-cond-2}).

Lastly, we estimate the attraction between two different blobs. 
Again, we assume two blobs align along $z$-coordinate so that the peak of blob A is at $z=\ell_A$ and that of blob B is at $z=-\ell_B$.
The neck waist is also assumed at $z=0$. The distance between blobs are given by
\begin{equation}
L = \ell_A+\ell_B.\label{eq:kin-L-ell}
\end{equation}
Away from the waist, we estimate the mass profile with a naive superposition
\begin{equation}
 m \simeq m_A e^{-\fr{2}(z-\ell_A)^2}e^{-\frac{|{\bf y}|^2}{2}}+  m_B e^{-\fr{2}(z+\ell_B)^2}e^{-\frac{|{\bf y}|^2}{2}},\label{eq:kin-twoblob-estimate}
\end{equation}
where ${\bf y}$ is the perpendicular directions. Each blob mass is given by
\begin{equation}
\cM_A\simeq (2\pi)^{\frac{p}{2}}m_A,\quad \cM_B\simeq (2\pi)^{\frac{p}{2}}m_B.
\end{equation}
Now, we determine the distance between the peaks and waist.
The waist should appear where the two Gaussians cross
\begin{equation}
 m_A e^{-\frac{\ell_A^2}{2}} \simeq m_B e^{-\frac{\ell_B^2}{2}}.
\end{equation}
Combining with eq.~(\ref{eq:kin-L-ell}), we obtain
\begin{equation}
\ell_A \simeq \frac{L}{2} + \fr{L}\log \frac{m_A}{m_B},\quad \ell_B \simeq \frac{L}{2} - \fr{L}\log \frac{m_A}{m_B}.
\end{equation}
This shows that the mass difference only has the sub-leading effect, and hence we expect the black string result can be applied to the leading order estimate.
Expanding eq.~(\ref{eq:kin-twoblob-estimate}) with $z=\ord{1}$ and $L\gg1$, we obtain
\begin{equation}
m\simeq \sqrt{m_A m_B} e^{-\frac{L^2}{8}} \left(e^{\frac{Lz}{2}}+ e^{-\frac{Lz}{2}}\right)e^{-\frac{z^2}{2}}.
\end{equation}
This cannot be used directly on the waist. Instead, this should be considered as the boundary condition for the neck solution. Repeating the analysis in section~\ref{sec:nubs-neck-waist}, we obtain the one dimensional tension
\begin{equation}
 \tau \simeq \frac{\sqrt{m_A m_B}}{2}L^2 e^{-\frac{L^2}{8}}.
\end{equation}
Thus, the attraction force~(\ref{eq:kin-equal-centralf}) is generalized to
\begin{equation}
F(r) = \frac{\sqrt{\cM_A \cM_B} \, r^2}{2\sqrt{2\pi}} e^{-\frac{r^2}{8}}.
\end{equation}
This contrasts with Newton's law which depends linearly on both masses.

\section{Conclusion}\label{sec:summary}
In this article we have developed an analytical treatment of the blob-blob interaction in the large $D$ effective theory, in which the motion of blobs is approximately described by Newtonian mechanics.
We have discovered that the brane tension on the thin neck between blobs plays a key role in generating the attractive force. 

We started by revisiting the black string analysis for small tension, 
where non-uniform black strings are solved by expanding from the zero tension Gaussian solution.
The scaling hierarchy between the neck waist and compactification length $L$ enabled the matched asymptotic expansion between the narrow waist and blob tail.\footnote{For $L\simeq \ords{\sqrt{D}}$, this scaling hierarchy would smoothly continue to the near-neck structure in the fused conifold~\cite{Emparan:2019obu}.} This hierarchy was also kept in the physical quantities as a trans-series with respect to $L$.

In rigidly rotating cases, the black dumbbells and ripples found in ref.~\cite{Licht:2020odx}
have been also studied by using the interactive blob approximation.
The logarithmically extending necks in the limit $\Omega \to 0$, observed in the numerical analysis,
have been analytically confirmed for both solutions as the result of the balance between the centrifugal force and brane tension. In the ripples, the long neck whose length grows as $\Omega^{-1}$ has been also resolved in the same way. 

We have shown that in the slowly rotating limit $\Omega = \ords{1/\sqrt{D}}$, the thermodynamics of the single blob ring is equivalent to that of large $D$ black rings that were constructed in the effective theory using ring-adapted coordinates~\cite{Tanabe:2015hda}. This implies that blob rings can be identified as part of the large $D$ black rings/Saturns solved in the ring coordinate.
Beyond the regime of slow rotation, however,
it remains unclear whether ripples are actually ripples or rings/Saturns, since the leading order theory cannot resolve the difference between them.

The large $D$ results have also been shown to be consistent with the large $D$ limit of the blackfold approach~\cite{Emparan:2007wm,Armas:2014bia}. 
Remarkably, we confirmed that the elastic bending effect on the blackfold coincides with the correction perturbative in $\Omega$ for the blob ring. 
This implies that the blob approximation is the large $D$ equivalent of the blackfold approximation.

Finally, we have formulated the mechanics of interacting blobs.
We have derived an equation of motion with the effective attraction force. 
The attraction comes from the boundary integrals around blobs, in which thinly connecting necks provide the dominant contribution. 
This effective equation of motion has allowed us to explain kinematic aspects of blob collisions observed in the numerical simulations of Ref.~\cite{Andrade:2019edf,Andrade:2018yqu,Andrade:2020ilm}, as well as to understand the balance of forces in highly-deformed dumbbells. 

The current analysis 
has been done for specific configurations and a more general and comprehensive formulation would be desired.
The effective Newtonian description pursued in section~\ref{sec:kinematic} is an important clue in this direction. Some simplifications like fixed boundaries and no boundary fluxes 
 should be reconsidered to make a more general effective theory.
We also ignored the boundary viscosity in the tension, which was important for the entropy production during the collision~\cite{Andrade:2020ilm}.
For the motion of extended blob objects like blob rings, one will need further extensions which can describe both extrinsic and intrinsic motions as in the blackfold approach.

In the leading order theory, the refined blob approximation guarantees that
the neck can extend arbitrary long as it gets thinner.
However, we expect that the $1/D$ expansion breaks down for very thin neck.
Eventually, this will lead to a pinch and a topology-changing transition.
In~ref.\cite{Emparan:2018bmi}, by numerically solving $4$NLO effective equation, the zero tension condition admitted $D^{1/4}$ dependence on the maximum period for the non-uniform black string. This scaling is a little smaller than $L = \ords{\sqrt{D}}$ in which the topology-changing transition was observed in the large $D$ conifold ansatz~\cite{Emparan:2019obu}.
The blob approximation with $1/D$ correction should solve this puzzle.

The phase of black ripples includes both actual black ripples and black rings/Saturns.
At $D=\infty$, since the horizon can be arbitrarily thin, the entire branch becomes a rippled horizon without a pinch, and rings/Saturns
exist only at $\Omega =0,\ {\cal J}/{\cal M}=\infty$.
As explained above, the leading order theory cannot see the transition from the ripples to the rings/Saturns. On the other hand, in the slow rotation limit $\Omega = \ords{1/\sqrt{D}}$, the blob ring result is shown to be consistent with the large $D$ result from the effective theory in the ring coordinate, in which the $1/D$ expansion breaks down around the long neck waist.
This implies that the transition should occurs somewhere for $\Omega > \ords{1/\sqrt{D}}$.
The $1/D$ correction should be taken into account to get further information on the transition.
Since the transition point goes to ${\cal J}/{\cal M}\to\infty$ for $D\to \infty$, at sufficiently large $D$, entire parts of black rings/Saturns would be treated as thin rings where the blob approximation is applicable, and there would be thin/fat ripples instead.\footnote{Here, the terms `thin' and `fat' are used regarding  the applicability of the blob approximation.} The critical value $\Omega_c$ would increase as the dimension gets lower. Finally, as was shown in $D=6,7$~\cite{Dias:2014cia}, the thin rings/Saturns take the place of thin ripples and a part of fat ripples turns to fat rings/Saturns.

One might use blob rings to study black ring dynamics at large $D$ as has been done for Myers-Perry black holes.
It is known that thin rings are unstable to Gregory-Laflamme (GL) type non-axisymmetric fluctuations~\cite{Santos:2015iua,Figueras:2015hkb,Armas:2019iqs}. In the large $D$ limit, the same instability was found in the effective theory on the ring coordinate~\cite{Tanabe:2015hda}. Surprisingly, by numerically evolving the ring effective equation~\cite{Chen:2018vbv}, in the case of the large $D$ black ring, the end point of the instability was shown to be a stable non-uniform black ring.
The dynamical study of blob rings should help understanding their dynamics beyond the slowly rotating regime.

The elastic instability numerically found in $D=5$~\cite{Figueras:2015hkb} is another interesting phenomenon, which deforms the ring keeping its thickness almost unchanged, in contrast to the GL instability.
Recently, it was shown that thin black rings are free of this type of instability within the scope of the blackfold approximation~\cite{Armas:2019iqs}.\footnote{The elastic instability found in the large $D$ ring~\cite{Tanabe:2016opw} was questioned in ref.~\cite{Armas:2019iqs}.}
It is interesting to see if blob rings admit the elastic-type instability with or without the non-perturbative correction caused by the blob-blob interaction.

As a straightforword application, one can construct black flowers~\cite{Licht:2020odx} or black brane lattices~\cite{Rozali:2016yhw,Dias:2017coo} in their highly deformed regime. By balancing the neck tension~(\ref{eq:kin-force-I}) with centrifugal forces, the multiple rotating blob configuration can be constructed for black flowers. 
In the brane lattice setup, the neck tension from multiple directions would rise the multipole perturbation around blobs.

It should be possible to extend the analysis developed here to solutions with Maxwell charge.
In the rigidly rotating case, the effective equation was shown to coincide with the neutral one with only a simple redefinition of variables~\cite{Andrade:2018rcx}.
It will be interesting to see how the charge repulsion ( attraction ) competes with the neck tension.
In the presence of a cosmological constant, as long as the blob solution exists, the same technique should apply. 
Note that the effective theory requires have the negative pressure in order to have localized, stationary blob solutions. 
The negative pressure in the effective theory originates from the curvature of the large dimensional sphere along the horizon; therefore the blob method does not apply to the effective theory of planar AdS black holes.

\section*{Acknowledgments}
The author would like to thank Roberto Emparan for useful comments and discussion.
The author also thanks Raimon Luna and David Licht for the discussion in the early stage of this work.
This work is supported by ERC Advanced Grant GravBHs-692951 and MEC grant FPA2016-76005-C2-2-P and JSPS KAKENHI Grant Number JP18K13541 and partly by Osaka City University Advanced Mathematical Institute (MEXT Joint Usage/Research Center on Mathematics and Theoretical Physics).


\appendix

\section{Detailed calculations in the black string analysis}\label{sec:neck-sub}
\subsection{Sub-leading correction in the neck analysis}
The next-to-leading order solution to eq.~(\ref{eq:nubs-effeq-neck}) is obtained straightforwardly,
\begin{align}
\bar{\cR}_1(\xi) =\bar{r}_1+16 \log \cosh(\xi/2) -2\left[\xi^2+4\xi+\pi^2/3+4\, {\rm Li}_2\left(-e^{-\xi}\right) \right]\tanh(\xi/2),
\end{align}
where we used the matching result~(\ref{eq:nubs-match-r0}).
Again the integration constant $\bar{r}_1$ is set for the neck minimum at $\xi=0$, which is determined by the asymptotic behavior
\begin{align}
 \bar{\cR}_1(\xi) \simeq -2\xi^2+\bar{r}_1-\frac{2\pi^2}{3}-16\log 2 + \ord{e^{-|\xi|}}.
 \end{align}
The match with the Gaussian blob~(\ref{eq:nubs-gaussian-0}) gives
\begin{equation}
 \bar{r}_1 =\frac{2\pi^2}{3}+ 16 \log2.
\end{equation}
The next-to-next-to-leading order solution $\cR_2(\xi)$ is obtained in the similar way.
The result is put in the appendix~\ref{sec:neck-NNLO}. Here, we only show the matching result for the neck minimum,
\begin{equation}
  \bar{r}_2 = \frac{64\pi^2}{3}+96 \zeta(3)+128 (\log^22+\log2).
\end{equation}
This leads to eqs.~(\ref{eq:Rmin-L-NLO}) and (\ref{eq:tension-L-NLO}).

\subsection{Neck solution at Next-to-next-to-leading order}\label{sec:neck-NNLO}
\begin{align}
&\bar{\cR}_2(\xi) = {\bar r}_2-96 \zeta
   (3)+\frac{64 \pi ^2}{3}-128 (\log ^2 2+ \log 2)+\frac{4 \left(432 \zeta (3)-96 \pi ^2+\pi ^4\right)}{9 X}-\frac{4 \pi ^4}{9 X^2}\nonum
&\quad+\frac{4(X-1)}{X^2} \log ^4(X-1)+\frac{32(1-X)}{X^2}\log ^3(X-1)+\left(-\frac{8 \left(24+\pi ^2\right)}{3
   X^2}+\frac{8 \pi ^2}{3 X}+64\right) \log ^2(X-1)\nonum
&\quad+\left(\frac{32 \pi ^2}{3 X^2}-\frac{32 \left(\pi
   ^2-12\right)}{3 X}-128\right) \log (X-1)+\frac{32(2-X)}{X} \log ^3 X\nonum
 & \quad 
 +\left(\frac{96(X-2)}{X} \log (X-1)+128\right) \log ^2 X +\left(128+\frac{80 \pi   ^2}{3}-\frac{160 \pi ^2}{3  X}\right) \log X\nonum
 &\quad +\left(\frac{16(2-X)}{X} \log ^2(X-1)-\frac{64(X+2)}{X} \log (X-1)\right) \log X+\frac{64(X-1)}{X^2}\text{Li}_2(1-X){}^2\nonum
&\quad+\text{Li}_2(1-X) \left(256+\frac{32 \left(\pi ^2-48\right)}{3   X}-\frac{32 \pi ^2}{3 X^2}+\frac{32(X-1)}{X^2} \log   ^2(X-1)+\frac{128(X-2)}{X} \log X\right.\nonum
&\left. \quad \qquad -\frac{64(X^2-2)}{X^2}\log (X-1)\right)+\frac{96(X-2)}{X}\left(  \text{Li}_3(1-X)+2\text{Li}_3\left(\frac{1}{X}\right)\right).
\end{align}
where $X:=e^{-\xi}+1$. The integration constant corresponds to the value at $\xi=0$.
\begin{equation}
\bar{\cal R}_2(0)=\bar{r}_2.
\end{equation}
For $|\xi| \to \infty$, the solution is expanded by
\begin{equation}
\bar{\cR}_2(\xi) \simeq \bar{r}_2  - \frac{64\pi^2}{3} -96 \zeta(3) -128(\log 2+\log^2 2)+\ord{e^{-|\xi|}}.
\end{equation}

\subsection{Total mass up to sub-leading orders}\label{sec:neck-sub-mass}
Expanding the blob part~(\ref{eq:nubs-mass-blob-exact}) up to one higher order, we obtain
\begin{align}
& \frac{{\cal M}_{\rm blob}}{{\cal M}_0} = 1 - \frac{2 L e^{-\frac{L^2}{8}}}{\sqrt{2\pi}} \left[1+\frac{2}{L^2}\left(\frac{\pi^2}{3}+2+2\xi_c+e^{\xi_c}\right)+ \frac{4}{L^4}\left( 24 \zeta (3)+12+\frac{14 \pi ^2}{3}+\frac{ \pi ^4}{18}\right.\right.\nonum
 &\left. \left. \hspace{2.5cm}+\,\frac{2}{3} \left(18+\pi ^2\right) \xi_c+4 \xi_c^2-e^{\xi_c} \left( \xi_c^2-2 \xi_c+2\right)\right)+\ord{\tau,e^{-\xi_c},L^{-6}}\right].
\end{align}
By using the neck solution up to NLO in the neck part~(\ref{eq:nubs-m-neck-LO}), the expansion becomes
\begin{align}
& \frac{{\cal M}_{\rm neck}} {{\cal M}_0}
=\frac{4 e^{-\frac{L^2}{8}}}{\sqrt{2\pi} L} \left(2\xi_c + e^{\xi_c}+\frac{2}{L^2}\left(16+\frac{4\pi^2}{3}+\frac{2}{3}(18+\pi^2)\xi_c + 4 \xi_c^2 - e^{\xi_c}(\xi_c^2-2\xi_c+2)\right)+\ord{e^{-\xi_c},L^{-4}}\right).\label{eq:nubs-m-neck}
\end{align}
Note that the neck part now has a finite contribution in the sub-leading order, not just the cut off counter terms. The summation gives the cut off independent result
\begin{equation}
\frac{{\cal M}}{{\cal M}_0} = 1  - \frac{2 L e^{-\frac{L^2}{8}}}{\sqrt{2\pi}} \left[1+\frac{4}{L^2}\left(\frac{\pi^2}{6}+1\right)+\frac{2(432\zeta(3)+\pi^4+60\pi^2-72)}{9L^4}+\ord{L^{-6}}\right].\label{eq:bs-mass-res}
\end{equation}

\section{Calculations for ripples}
\subsection{Single blob ring}\label{sec:app-ripple-single}
The absence of the short neck simplifies the analysis of the single blob ring in $N=2$ and $N=3$ ripples. Ignoring the non-perturbative correction, the solution is found by simply expanding in $\Omega$,
\begin{align}
& \cR(z) = \fr{2}-\frac{z^2}{2}+\frac{3}{2}\Omega^2\left(z^2-2\right)
-  \fr{2}\Omega^3z^3+ \Omega^4\left(\frac{z^4}{3}+4z^2-\frac{19}{2}\right)
- \Omega^5 \left(\frac{z^5}{4}+\frac{11z^3}{6}\right)\nonum
&+\Omega ^6
   \left(\frac{z^6}{5}+\frac{97 z^4}{72}+\frac{71 z^2}{3}-\frac{805}{12}\right)-\Omega ^7 \left(\frac{z^7}{6}+\frac{67 z^5}{60}+\frac{247 z^3}{18}\right)\nonum
   &+\Omega ^8 \left(\frac{z^8}{7}+\frac{1769 z^6}{1800}+\frac{6137 z^4}{540}+\frac{34853
   z^2}{180}-\frac{231767}{360}\right)-\Omega ^9 \left(\frac{z^9}{8}+\frac{2257 z^7}{2520}+\frac{37723 z^5}{3600}+\frac{74213 z^3}{540}\right)\nonum
  &+\Omega ^{10} \left(\frac{z^{10}}{9}+\frac{6543 z^8}{7840}+\frac{1921093 z^6}{189000}+\frac{29460041
   z^4}{226800}+\frac{74689297 z^2}{37800}-\frac{570064693}{75600}\right),
\label{eq:app-ripple-single-ring}
\end{align}
where $z:=r-L$ and we show the terms up to $\ord{\Omega^{10}}$. The blob peak is determined by
\begin{equation}
 L = \Omega^{-1} + \frac{3}{2}\Omega^3+\frac{79}{4}\Omega^5 
+ \frac{3065}{12} \Omega^7 + \frac{2562551}{720} \Omega^9+\ord{\Omega^{11}}.
\label{eq:app-ripple-single-L-pert}
\end{equation}
The integration constant is fixed so that the balance condition holds up to the relevant order
\begin{equation}
 \int^\infty_{-\infty} dz \left(- \frac{\cR'(z)^2}{L+z}+\Omega^2 (L+z)\right) e^{\cR(z)} = 0.
\end{equation}

\subsection{$\ord{\Omega^2}$-correction to blob rings}
Up to $\ord{\Omega}$, the $i$-th blob ring solution is expressed by
\begin{equation}
 \cR^{(i)}_{0}(z_i) = 1-\frac{w_i^2}{2} -\frac{z_i^2}{2},
\end{equation}
where
\begin{equation}
z_i:=r-L_i,\quad  w_i := \Omega L_i = 1 + \delta^{[1]}\! L_i \Omega +\delta^{[2]}\! L_i \Omega^2+\dots
\end{equation}
The $\ord{\Omega^2}$ correction $\cR^{(i)}(z_i) = \cR^{(i)}_0(z_i)+\Omega^2 \delta \cR^{(i)}(z_i)$ is solved by
\begin{align}
& \delta \cR^{(i)}(z) = 
-3+\frac{3z^2}{2}-2 \delta^{[1]} \! L_i z 
- \delta^{[1]} \! L_i\, z^3 \hgfunc{2}{2} \left(1,1;\fr{2},2;-\frac{z^2}{2}\right) +\sqrt{2\pi} \delta^{[1]} \! L_i \, z^2 e^{-\frac{z^2}{2}} {\rm erfi}\left(\frac{z}{\sqrt{2}}\right)\nonum
&
\qquad + \left(\cC_i+\sqrt{2\pi}\delta^{[1]} \! L_i \, {\rm erf} \left(\frac{z}{\sqrt{2}}\right)\right)\left(e^{\frac{z^2}{2}} -\sqrt{\frac{\pi}{2}} z\, {\rm erfi}\left(\frac{z}{\sqrt{2}}\right)\right) .\label{eq:app-ripple-dR}
\end{align}
where $\cC_i$ is the integration constant which modifies the peak maximum by $\delta \cR^{(i)}(0) = \cC_i-3$. The limit $z \to \pm \infty$ admits the exponential growth,
\begin{equation}
 \delta \cR^{(i)}(z) \simeq z^{-2} e^{\frac{z^2}{2}}\left(\cC_i \mp\sqrt{2\pi} \delta^{[1]}\! L_i\right).
 \label{eq:app-ripple-dR-as}
\end{equation}
To determine the integration constant, we evaluate the integral~(\ref{eq:ripple-tension-integral}) from the neck waist $r=L_i-\Delta_{i-1}/2$ to the peak $r=L_i$,\footnote{For $i=1$, the same result is obtained by replacing the integral domain with $[L_i,L_i+\Delta_i/2]$.}
\begin{equation}
\left[ \left(\fr{2} \cR'^2+\cR-1+\fr{2}\Omega^2 r^2\right)e^\cR \right]^{L_i}_{L_i-\frac{\Delta_{i-1}}{2}}
= \sqrt{e} \Omega^2 \left(\cC_i-3+2\sqrt{2\pi} \sum_{j=i}^s \delta^{[1]}\!L_i\right),
\end{equation}
where the neck tension in eq.~(\ref{eq:ripple-tension-rhs}) is used.
The left hand side integral is easily evaluated with $\cR_0^{(i)}(z)$
\begin{equation}
 \int^{L_i}_{L_i-\frac{\Delta_{i-1}}{2}} dr \left(-\fr{r}(\cR')^2+\Omega^2 r\right) e^\cR 
 = \sqrt{e} \Omega^2 \left(-3+\sqrt{2\pi} \delta^{[1]} \! L_i\right)+\ord{\Omega^3}.
\end{equation}
Therefore, we obtain
\begin{equation}
\cC_i = - 2 \sqrt{2\pi}\sum_{j=i+1}^s \delta^{[1]}\!L_i-\sqrt{2\pi} \delta^{[1]} \! L_i.
\label{eq:app-ripple-dR-coef-res}
\end{equation}
By substituting this to eq.~(\ref{eq:app-ripple-dR-as}),
one can see that the innermost and outermost blog rings are absent of the exponential growth in the inward or outward direction, which enables the connection to the long neck or asymptotic region.

\subsection{Evaluation of $\langle \delta^{[2]}\! L \rangle$}
\label{app:ripple-dL2}
To calculated the average of $\ord{\Omega^2}$-correction to $L_i$, we evaluate the integral on the right hand side of eq.~(\ref{eq:ripple-tension-integral}) from the long neck waist to the infinity up to $\ord{\Omega^3}$. First, we focus on the integral over $i$-th blob ring
\begin{align}
& \int_{i{\rm \mathchar`-th}} dr \left( - \fr{r} \cR'(r)^2+\Omega^2 r\right)e^{\cR(r)} \simeq 
 \sqrt{2\pi} e \Omega^2 ( 2 \delta^{[1]} L_i+ (2\delta^{[2]} L_i -  3(\delta^{[1]} L_i)^2-3)\Omega) \nonum
 &\hspace{1.2cm} +\Omega \left[z_i e^{-\frac{z_i^2}{2}} \right]^{z_{i,+}}_{z_{i,-}} + \Omega^3 \int^{z_{i,+}}_{z_{i,-}} dz_i \left(2 z_i \delta \cR^{(i)}{}'(z_i)+(1-z_i^2) \delta \cR^{(i)}(z_i)\right) e^{1-\frac{z_i^2}{2}}
\end{align}
where $z_i := r-L_i$ and the cut off is given by 
\begin{equation}
z_{i,+}= \frac{\Delta_i}{2} + \frac{2}{\Delta_i} \xi_c,\quad z_{i,-}= -\frac{\Delta_{i-1}}{2} + \frac{2}{\Delta_{i-1}} \xi'_c.
\end{equation}
As in the black string analysis, we expect the cut off dependence will cancel out
together with the neck contribution. Here, instead of evaluating the neck contribution, we simply neglect the cut off dependence. One should note that the neck integral can also provide the finite contribution at higher order in $1/\Delta$ as in eq.~(\ref{eq:nubs-m-neck}).
The integral involving $\delta \cR^{(i)}(z_i)$ is evaluated by using eqs.~(\ref{eq:app-ripple-dR})
and (\ref{eq:app-ripple-dR-coef-res}),
\begin{align}
&\int^{z_{i,+}}_{z_{i,-}} dz_i \left(2 z_i \delta \cR^{(i)}{}'(z_i)+(1-z_i^2) \delta \cR^{(i)}(z_i)\right) e^{1-\frac{z_i^2}{2}}\nonum
&= 3\sqrt{2\pi}e +\sqrt{2\pi} e\left( \sum_{j=i+1}^s \delta^{[1]}\!L_j \Delta_i +\sum_{j=i}^s \delta^{[1]}\!L_j \Delta_{i-1}\right)+\ord{\Delta^{-2}}.
\end{align}
By summing the contribution from all blob rings, we obtain the condition
\begin{align}
 \sqrt{2\pi} \, e s \, \Omega^3 \left( 2\langle \delta ^{[2]} L \rangle- 3 \langle \delta^{[1]}L^2\rangle
 - \fr{s}\sum_{i=1}^{s-1} \sum_{j=i}^s \delta^{[1]} \! L_j \Delta_{i}\right) = \ord{\Omega^4},
\end{align}
where NLO result $\langle \delta^{[1]}\! L \rangle=0$ is already used.
Recalling the definition $\Delta_i:=\delta^{[1]} \! L_{i+1}-\delta^{[1]}\! L_{i}$, the resummation leads
to the identity,
\begin{equation}
 \fr{s} \sum_{i=1}^{s-1} \sum_{j=i}^s \delta^{[1]} \! L_j \Delta_{i} =  \langle \delta^{[1]}\! L^2\rangle,
\end{equation}
with which eq.~(\ref{eq:ripple-ave-dL2-res}) is obtained.

\bibliographystyle{JHEP}
\bibliography{BlobsNecks}

\end{document}